\documentclass[%
 aip,
 amsmath,amssymb,
 reprint,%
]{revtex4-1}

\usepackage{graphicx}
\usepackage{dcolumn}
\usepackage{bm}

\usepackage[utf8]{inputenc}
\usepackage[T1]{fontenc}
\usepackage{mathptmx}

\newcolumntype{d}[1]{D{.}{.}{#1}}
\usepackage{color}
\usepackage{xspace}

\graphicspath{{./graphs/}}

\newcommand{\WignerD}[3]{\ensuremath{D^{#1}_{#2}({#3})}}

\def\Inv/{\hat{I}}
\def\Unity/{\hat{E}}

\def\Htwoatsixty/{H$_2$@C$_{60}$}
\def\Htwoatseventy/{H$_2$@C$_{70}$}
\def\HDatsixty/{HD@C$_{60}$}
\def\Dtwoatsixty/{D$_2$@C$_{60}$}
\def\wateratsixty/{\ensuremath{\mathrm{H}_2\mathrm{O}@\mathrm{C}_{60}}}
\def\water/{\ensuremath{\mathrm{H}_2\mathrm{O}}}
\def\HFatsixty/{HF@C$_{60}$}
\def\nhopen/{NH$_3$@ocf1}
\def\chopen/{CH$_4$@ocf2}
\def\ch4/{CH$_4$}
\def\ch4c60/{CH$_4$@C$_{60}$}
\def\nh3/{NH$_3$}
\def\waterc60/{H$_2$O@C$_{60}$}
\def\wateropen/{H$_2$O@ocf}
\def\wateropenOne/{H$_2$O@ocf1}
\def\htwoatsixty/{H$_2$@C$_{60}$}
\def\hfatsixty/{HF@C$_{60}$}

\def\Csixty/{\ensuremath{\mathrm{C}_{60}}}
\def\Cseventy/{C$_{70}$}
\def\Htwo/{H$_2$}
\def\Dtwo/{D$_2$}
\def\para/{{\it para\/}}
\def\ortho/{{\it ortho\/}}
\def\wn/{\ensuremath{\,\mathrm{cm}^{-1}}}
\def\area/{\,cm$^{-2}$}
\def\sikma/{{\ensuremath\,\mathrm{cm}/\mathrm{molecule}}}

\def\svec/{\mathbf{s}}

\def\rivec/{\mathbf{r}_i}
\def\ri/{r_i}
\def\thetai/{\theta_i}
\def\phii/{\phi_i}
\def\riunitvec/{\hat{\mathbf r}_i}
\def\thetaiunitvec/{\hat{\pmb\theta}_i}
\def\phiiunitvec/{\hat{\pmb\phi}_i}

\def\RCMvec/{\mathbf{R}}
\def\RCvec/{\mathbf{R}_G}
\def\RCM/{R}
\def\thetaCM/{\Theta}
\def\phiCM/{\Phi}

\def\CMshift/{\mathbf{x}}

\def\RCivec/{\mathbf{R}_i}
\def\RCi/{R_i}
\def\thetaCi/{\Theta_i}
\def\phiCi/{\Phi_i}
\def\RCiunitvec/{\hat{\mathbf R}_i}
\def\thetaCiunitvec/{\hat{\pmb\Theta}_i}
\def\phiCiunitvec/{\hat{\pmb\Phi}_i}

\def\angulari/{\Omega_{i}}
\def\angularCM/{\Omega}
\def\angularCi/{\Omega_{Ci}}
\def\angularHtwo/{\Omega_s}

\def\ket#1{\ensuremath{\left| #1 \right>}}
\def\bra#1{\ensuremath{\left< #1 \right|}}

\def\vv{\mathbf{v}}

\def\chisq/{\chi^2}
\def\chisqmin/{\chi^2_{\mathrm{min}}}
\def\parvec/{\mathbf{a}}
\def\parvecmin/{\mathbf{a}_{\mathrm{min}}}
\newcommand{\matx}[1]{\bar{\bm{#1}}}
\newcommand{\vect}[1]{\mathbf{#1}}

\begin{document}

\title{Infrared spectroscopy of endohedral  \water/ in \Csixty/.}

\author{A. Shugai}
\author{U. Nagel}
\affiliation{National Institute of Chemical Physics and Biophysics, Akadeemia tee 23, 12618 Tallinn, Estonia}
\author{Y. Murata}
\affiliation{Institute for Chemical Research, Kyoto University, Kyoto 611-0011, Japan}
\author{Yongjun Li}
\affiliation{Department of Chemistry, Columbia University, New York, New York 10027,USA}
 \altaffiliation[Present address ]{Merck \& Co., 126 E Lincoln Ave, Rahway, New Jersey, 07065, USA.}
\author{S. Mamone}
\author{A. Krachmalnicoff}
\author{S. Alom}
\author{R. J. Whitby}
\author{M. H. Levitt}
\affiliation{School of Chemistry, Southampton University, Southampton SO17 1BJ, United Kingdom}
\author{T. R{\~o}{\~o}m}
\email{toomas.room@kbfi.ee}
\affiliation{National Institute of Chemical Physics and Biophysics, Akadeemia tee 23, 12618 Tallinn, Estonia}

\date{\today}

\begin{abstract}

Infrared absorption  spectroscopy study of  endohedral water molecule in a solid mixture of \wateratsixty/ and \Csixty/ was carried out   at liquid helium temperature.
From the  evolution of the spectra during the \ortho/-\para/ conversion process, the spectral lines were identified as \para/- and \ortho/-\water/ transitions.  
Eight vibrational transitions with rotational side peaks were observed in the mid-infrared: $\omega_1$, $\omega_2$,  $\omega_3$, $2\omega_1$, $2\omega_2$, $\omega_1 +\omega_3$, $\omega_2 +\omega_3$,  and   $2\omega_2+\omega_3$. 
The vibrational frequencies $\omega_2$ and 2$\omega_2$ are lower by 1.6\% and the rest by 2.4\%,  as compared to  free \water/.
A model consisting of a rovibrational Hamiltonian with the dipole and quadrupole moments of \water/ interacting with the crystal field    was used to fit the infrared absorption  spectra.
The electric quadrupole interaction with the crystal field lifts the degeneracy of the rotational levels. 
The finite amplitudes of the pure $v_1$ and  $v_2$ vibrational transitions are consistent with the interaction of the water molecule dipole moment with a lattice-induced electric field. 
The permanent dipole moment of encapsulated \water/ is found to be  $0.5\pm 0.1$ D as determined from  the far-infrared rotational line intensities.
The  translational mode of the  quantized center of mass motion of \water/ in the molecular cage of \Csixty/ was observed at 110\wn/ (13.6\,meV). 

\end{abstract}


\maketitle

\section{Introduction}
Endohedral fullerenes consist of atoms or molecules  fully encapsulated in closed carbon cages. 
The remarkable synthetic route known as ``molecular surgery" has led to the synthesis of atomic endofullerenes He@C$_{60}$~\cite{Morinaka2010} and Ar@C$_{60}$~\cite{Bloodworth2020Ar}, and  several molecular endofullerene species, including \Htwoatsixty/~\cite{Komatsu2005}, \wateratsixty/~\cite{Kurotobi2011},  \HFatsixty/~\cite{Krachmalnicoff2016}, \ch4c60/~\cite{Bloodworth2019},  and their isotopologs \cite{Horsewill2010,Ge2011D2HD,Krachmalnicoff2014}. 
It is well established that these endohedral molecules do not form chemical bonds with the carbon cage and rotate freely~\cite{Krachmalnicoff2016,Mamone2009,Beduz2012}.
The rotation is  further facilitated by the nearly spherical  symmetry of the \Csixty/ cage.
The thermal and chemical stability of A@C$_{60}$ opens up the unique possibility of studying the dynamics and the interactions of a small molecule with carbon nano-surfaces.
The trapping potential of dihydrogen \Htwo/ has been described with high accuracy using infrared (IR) spectroscopy~\cite{Ge2011,Ge2011D2HD}, inelastic neutron scattering (INS)~\cite{Horsewill2012} and theoretical calculations~\cite{Xu2009}.
The non-spherical shape of the small molecule and the quantized translational motion of its center of mass leads to coupled rotational and translational dynamics~\cite{Xu2008,Mamone2009,Felker2016}.

The water-endofullerene \wateratsixty/ is of particular interest. The encapsulated water molecule possesses rich spatial quantum dynamics. It is an asymmetric-top rotor,   supports three vibrational modes,  displays nuclear spin isomerism (\para/- and \ortho/-water) and has an electric dipole and quadrupole moment.

Low-temperature dielectric measurements \cite{Meier2015} on solid \wateratsixty/ show that the  electric dipole moment of the encapsulated water is reduced to $0.51\pm0.05$\,D  from  the free water value 1.85\,D. 
The \Csixty/ carbon cage responds to the endohedral  water molecule with a  counteracting induced dipole, resulting in the lower total dipole moment\cite{Ensing2012}.

The dynamics of isolated or encapsulated single water molecules have been studied before in other environments, such as noble gas matrices, solid hydrogen and liquid helium droplets. Although the trapping sites in these matrices have high symmetry and allow water rotation, these systems exist only at low temperature~\cite{Ceponkus2013,Fajardo2004}
or for a very short time~\cite{Lindsay2006,Kuyanov2006}. 
Water has also been studied in crystalline environments with nano-size cavities.
However, in this case, the interactions with the trapping sites inhibit the free rotation of the water molecules \cite{Gorshunov2013, Zhukova2014, Kolesnikov2016,Belyanchikov2020}. 

Several spectroscopic techniques have been used to study \wateratsixty/,  including nuclear magnetic resonance (NMR), INS and IR~\cite{Beduz2012,Mamone2011CCR, Meier2018, Kurotobi2011, Li2012} and time-domain THz spectroscopy \cite{Zhukov2020}.
The low-lying rotational states of the encapsulated molecule are found to be very similar to those of an isolated water molecule, with the notable exception of a 0.6\,meV splitting in the $J=1$ rotational state~\cite{Beduz2012,Goh2014}. 
This indicates that the local environment of the water molecule in \wateratsixty/ has a lower symmetry than the icosahedral point group of the encapsulating \Csixty/ cage. 
The splitting has been attributed to the interaction between the electric quadrupole moment of \water/ and the electric field gradients generated by the electronic charge distribution of neighbouring \Csixty/ molecules \cite{Felker2017,Bacic2018FD}.
The merohedral disorder present in solid \Csixty/ leads to two \Csixty/ sites with different quadrupolar interactions. This merohedral disorder also leads to splittings of the IR phonons in solid \Csixty/ ~\cite{Homes1994}. 
There is also evidence from dielectric measurements that merohedral disorder leads to electric dipolar activity in solid \Csixty/~\cite{Alers1992}.

Confined molecules exhibit quantization of their translational motion (``particle in a box"), in addition to their quantized rotational and vibrational modes. 
Quantized translational modes have been observed at 60 to 70\wn/ for water in noble gas matrices~\cite{ AbouafMarguin2009,Ceponkus2013}. 
However, comparatively little is known about the centre-of-mass translational mode of \wateratsixty/. 
The fundamental frequency of the water translation mode in \wateratsixty/ has been predicted to occur at $\sim$160\wn/~\cite{Felker2016}. This relatively high frequency reflects the rather tight confinement of the water molecule in the \Csixty/ cage.

The energy level separation between the ground \para/ rotational state  and the lowest \ortho/ rotational state  in \wateratsixty/ has been determined to be 2.6\,meV (21\wn/) by INS \cite{Beduz2012}. 
This energy level separation corresponds to a temperature of 28\,K. 
The full thermal equilibration of \wateratsixty/ at temperatures below 30K therefore requires the conversion of \ortho/
into \para/ water. This conversion process takes between tens of minutes to several hours below 20K in \wateratsixty/ \cite{Beduz2012,Mamone2014,Meier2015,Zhukov2020}.
The spin-isomer conversion is much faster at ambient temperature, with a time constant of about 30\,s reported for \wateratsixty/ dissolved in toluene~\cite{Meier2018}.

In this paper, we report on a detailed low-temperature  far- and mid-IR spectroscopic study of \wateratsixty/ and \Csixty/ solid mixtures.
The IR technique allows us to measure the frequencies of rotational, vibrational and translational modes and from the line intensities to determine the dipole moment of encapsulated water.
In addition, the IR spectra reveal the interaction of endohedral water with the electrostatic fields present in solid \Csixty/.

The rest of this paper is organized as follows. 
Section \ref{sec:methods} discusses the sample preparation, the recording procedures of the IR spectra, and the determination of the IR absorption cross-sections for different filling factors, temperatures and \ortho/-\para/ ratios.
The quantum mechanical vibrating rotor model for the encapsulated water molecules is introduced in  Section \ref{sec:quantum_model}. 
We include in this model the interactions between the   \water/ electric dipole and quadrupole moments with the electrostatic fields present in solid \Csixty/. 
The theory of the IR line intensities is presented in Section \ref{sec:abs_cross_section}.
Section \ref{sec:results} presents the measured IR spectra and the fitting of this data by the quantum-mechanical model. The results are discussed in 
Section \ref{sec:discussion}, followed by a summary in Section \ref{sec:summary}.
The Appendix contains more detailed theory for the interaction of the water molecules with the electrostatic fields and the infrared radiation, and more details on the fitting of the experimental data by the quantum-mechanical model.

\section{Methods \label{sec:methods}}
\subsection{Sample preparation}

\wateratsixty/ was prepared  by multi-step synthetic route known as  ``molecular surgery"~\cite{Kurotobi2011,Krachmalnicoff2014}. 
The \water/-filled (number density  $N_\bullet$) and empty (number density  $N_\circ$) \Csixty/ were mixed and co-sublimed to produce small solvent-free crystals with a   filling factor $f=N_\bullet/(N_\bullet+N_\circ)$.
Five samples with filling factors  $f=0.014, 0.052, 0.10, 0.18$, and 0.80 were studied.
The powdered samples were pressed into pellets under vacuum.
The diameter of sample pellets was  3\,mm and the thickness $d$ varied from   0.2\,mm  to 2\,mm. 
The thinner samples were used in the mid-IR because of light scattering in the powder sample.
Samples were thicker for lower filling factors and thinner for higher filling factors to avoid the saturation of absorption lines in the far-IR.

\subsection{Measurement techniques}
\label{meaurement_tech}

The far-IR measurements  were done with a Martin-Puplett type interferometer and $^3$He cooled bolometer from 5 to 200\wn/ as described in Ref.~\onlinecite{Beduz2012}.
The IR measurements  between 600\wn/ and 12000\wn/ were performed with an  interferometer Vertex 80v (Bruker Optics) as described in Ref.~\onlinecite{Ge2011}.

Two methods were used to record the \wateratsixty/ absorption spectra.

{\em Method 1.}
The intensity through the sample, $I_s$, was referenced to the intensity through a 3\,mm diameter  hole, $I_0$.
The sample was allowed to reach \ortho/-\para/ thermal equilibrium at a temperature of 30 or 45\,K, and the temperature was rapidly reduced to 
10 or to 5\,K. 
The sample spectrum $I_s(t=0)$ was recorded immediately after the temperature jump. 
Since the \ortho/-\para/ conversion process is slow, the \ortho/ fraction  was assumed to be preserved during the $T$ jump, corresponding to the high temperature \ortho/ fraction  $n_{\mathrm o}\approx0.74$.
The absorption coefficient $\alpha$ was calculated from the ratio $Tr=I_s(t=0)/I_0$  as $\alpha(t=0) = -d^{-1} \ln[(1-R)^{-2}\,\, Tr ]$ 
where factor 
$(1-R)^{2} $ with $R=(\eta-1)^2(\eta+1)^{-2}$ 
corrects for the losses of radiation,  one  reflection from the sample front and one from the back face. 
The refraction index of solid \Csixty/ was assumed to be given by $\eta=2$~\cite{Homes1994}.
To identify \para/- and \ortho/-water absorption peaks the difference of two spectra was calculated, $\Delta \alpha = \alpha(t=0) - \alpha(\Delta t)$, where $\alpha(\Delta t)$ is the absorption spectrum measured after the waiting time   $\Delta t$.
Only the \para /- and \ortho/-\wateratsixty/ peaks show up  in the differential absorption  spectra, with the \para /- and \ortho/-\wateratsixty/ peak amplitudes having different signs, \ortho/ positive and \para/ negative.
This method was used for the far- and mid-IR part of the spectrum.

{\em Method 2.}
The sample was allowed to reach \ortho/-\para/ thermal equilibrium at a temperature of 30 or 45\,K, leading to an \ortho/-rich state,  as in the first method. 
The temperature was rapidly reduced to  10 or to 5\,K and a series of spectra recorded at intervals of a few minutes starting immediately after the $T$ jump, and continued until the \ortho/-\para/ equilibrium was reached. 
The differential absorption $\Delta\alpha = -d^{-1} \ln[I_s(0)/I_s]$ was calculated, where $I_s(0)$ is the spectrum recorded  immediately after the $T$ jump  and $I_s$ is the spectrum recorded when the low  temperature equilibrium was reached.
The equilibrium \ortho/ fraction is approximately $0.01$  at 5\,K. 
Method 2 was used  for the far-IR part of the spectrum.

\subsection{Line areas and absorption cross-sections}

The absorption line area $A_{ji}^{(k)}$ was determined by fitting the measured absorption $\alpha_{ji}^{(k)}(\omega)$ with Gaussian line shape.
$\ket{i}$ and $\ket{j}$ are the initial and final states of the transition and $k$ denotes \para/ ($k=\mathrm{p}$) or \ortho/ ($k=\mathrm{o}$) species.
From these experimental line areas, $A_{ji}^{(k)}$, a temperature and \para/ (\ortho/) fraction independent  line area $\overline{A_{ji}^{(k)}}$  was calculated,
\begin{eqnarray}\label{eq:line_area}
\overline{A_{ji}^{(k)}}(f)&=&   \frac{A_{ji}^{(k)}(f)}{  n_k (p_i^{(k)}-p_j^{(k)})}.
\end{eqnarray}
The population difference of initial and final states, $ p_i^{(k)}-p_f^{(k)}$, is given by the sample temperature $T$, while the \para/ (\ortho/) fraction $n_k$ depends on the history of the sample because of the \ortho/-\para/ conversion process.
The normalized absorption line area $\langle A_{ji}^{(k)}\rangle$ was determined from the linear fit of $\overline{A_{ji}^{(k)}}(f)$ for each absorption line,
\begin{equation}\label{eq:line_area_norm}
\overline{A_{ji}^{(k)}}(f)= f \langle A_{ji}^{(k)}\rangle.     
\end{equation}
Thus, the normalized absorption line area $\langle A_{ji}^{(k)}\rangle$ is the  absorption line area of a sample with a filling factor $f=1$   and spin isomer fraction  $n_k=1$   where all the population is in the initial state, $ p_i^{(k)}=1$.
We used $\langle A_{ji}^{(k)}\rangle$  to calculate a synthetic experimental spectrum  for the spectral fit with the quantum mechanical model, Section~\ref{sec:SpectralFitting}.

Furthermore, to compare the absorption cross-sections of \wateratsixty/ in solid \Csixty/ and  free \water/, a normalized absorption cross-section $\langle\sigma_{ji}^{(k)}\rangle$ was obtained as 
\begin{eqnarray}\label{eq:norm_crosssection}
\langle\sigma_{ji}^{(k)}\rangle &=& \left(\frac{3 \sqrt{\eta} }{  \eta^2 +2 } \right)^2 \frac{ \sigma_{ji}^{(k)}} {n_k ( p_i^{(k)}-p_j^{(k)})}\nonumber\\
&=& \left(\frac{3 \sqrt{\eta} }{  \eta^2 +2 } \right)^2 \frac{\langle A_{ji}^{(k)}\rangle} {N_{\mathrm C_{60}}},
\end{eqnarray}
where $\eta$ is the index of refraction of solid \Csixty/ and $\sigma_{ji}^{(k)}$ is the absorption cross-section of an endohedral water molecule,  Eq.~(\ref{eq:cross-section_op}) in Section~\ref{sec:abs_cross_section}.
This \wateratsixty/ absorption cross-section can be compared to the free water  normalized cross-section $\langle\sigma_{ji}^{(k)}\rangle = \sigma_{ji}^{(k)}n_k^{-1}( p_i^{(k)}-p_j^{(k)})^{-1}$, where  $p_i^{(k)}-p_j^{(k)}$ and  $n_k(T)$ are given by the temperature of the water vapour in the experiment reporting $\sigma_{ji}^{(k)}$.

\section{Theory \label{sec:theory}}

\begin{figure}
	\includegraphics[width=0.5\textwidth]{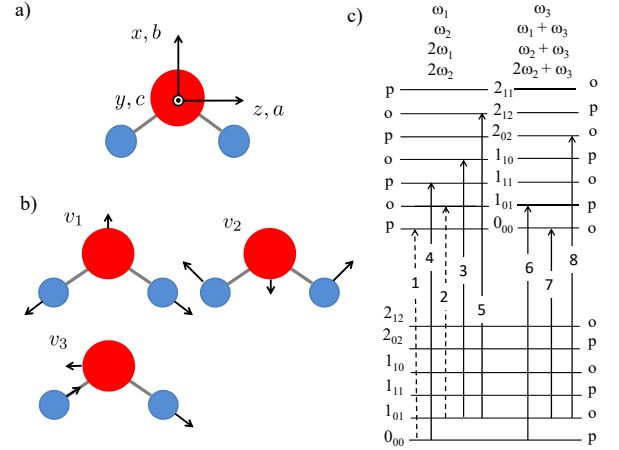}
	\caption{\label{fig:MolAxes}
		(a)	Molecule-fixed  coordinate frame $M=\{x,y,z\}$ and 
		the axes of principal moments of inertia, $\{a,b,c\}$.
		(b) Vibrations:  $v_1$ - symmetric stretch,  $v_2$ - symmetric bend and $v_3$ - asymmetric stretch.
		(c) \para/- and \ortho/-water rotational energy levels in the ground and excited vibrational states, see Section~\ref{sec:isomers}, and the rovibrational IR transitions (arrows) between the levels.
		The rotational states are labelled by  $J_{K_aK_c}$ and p (\para/) and o (\ortho/).
		 The IR transitions are between the \para/ or between the  \ortho/ states. 
		 1, 3, and 5 are \ortho/-\water/ transitions and 2, and 4 are \para/-\water/ transitions.
		 Transitions 1 and 2 are forbidden for a free water molecule.
		 The transitions where  the one-quantum excitation of the asymmetric stretch vibration $v_3$ is involved, are numbered 6 (\para/-\water/), and  7 and 8 (\ortho/-\water/ transitions).
		 The rotational and translational far-IR transitions in the ground vibrational states are shown in the inset to Fig.~\ref{fig:FIR_spectrum}.
	}
\end{figure}

\subsection{Quantum mechanical model of \wateratsixty/: Confined vibrating rotor in an electrostatic field \label{sec:quantum_model}}

We  use the following Hamiltonian  to model endohedral water molecule in solid \Csixty/:
\begin{equation}\label{eq:ham_HmVesHt}
\mathcal H = \mathcal H_{\mathrm M} + \mathcal H_{\mathrm {ES}} + \mathcal H_{\mathrm {T}},
\end{equation}
where $\mathcal H_{\mathrm M}=\mathcal H_{\mathrm{v}} + \mathcal H_{\mathrm {rot}} $ is the free-molecule rovibrational Hamiltonian and $\mathcal H_{\mathrm {ES}}$ is the electrostatic interaction of \water/ with the surrounding electric charges.
The translational Hamiltonian $\mathcal H_{\mathrm {T}}$  consists of water center of mass kinetic and potential energy in the molecular cavity of \Csixty/ molecule.

We neglect couplings between vibrational modes and between vibrational  and rotational modes.
Also, the coupling between translational motion and rotations is neglected.
In $\mathcal H_{\mathrm {ES}}$, terms describing the coupling of the solid \Csixty/ crystal field   to the electric dipole and quadrupole moment of \water/ are included.

The fitting of  IR absorption spectra, Section~\ref{sec:SpectralFitting}, is done with the Hamiltonian where the translational part is excluded:
\begin{equation}\label{eq:ham_HmVes}
\mathcal H = \mathcal H_{\mathrm M} + \mathcal H_{\mathrm {ES}}.
\end{equation}

We employ three coordinate frames.
The space-fixed coordinate frame is denoted $A$.
 $M=\{x,y,z\}$  is the molecule-fixed coordinate frame,  Fig.~\ref{fig:MolAxes}.
The Euler angles $\Omega_{A\rightarrow M}$ \cite{Zare1988,Bunker1998} transform $A$ to $M$.
The crystal coordinate frame is $C=\{x',y', z'\}$ with the $z'$  axis along the three-fold symmetry axis of the $S_6$ point group, the symmetry group of \Csixty/ site in solid \Csixty/. 
The Euler angles $\Omega_{A\rightarrow C}$ transform $A$ to $C$ and $\Omega_{C\rightarrow M}$ transform $C$ to $M$.
The  coordinate systems $A$ and $C$  are used because the radiation interacting with the molecule is defined in the space-fixed coordinate frame $A$ while the local electrostatic fields are defined by the crystal coordinate frame $C$, which in a powder sample has a uniform distribution of orientations relative to the  space-fixed frame $A$.

\subsubsection{Vibrations}

\water/ has three normal vibrations:  the symmetric stretch  of O--H bonds, quantum number $v_1$,  the  bending motion  of the H--O--H bond angle, $v_2$, and the asymmetric stretch of  O--H bonds, $v_3$ \cite{Bunker1998}, as sketched in Fig.~\ref{fig:MolAxes}(b).
The vibrational state is denoted $\ket{\vv}$ where the symbol $\vv$ denotes the three vibrational quantum numbers, $\vv\equiv v_1v_2v_3$,  each of which takes values $v_i\in\{0, 1, 2,\ldots\}$.
The vibrational energy for a harmonic vibrational potential is  
\begin{eqnarray}\label{eq:Evib}
E_{\vv} & = & \sum_{i=1}^3 \omega_i (v_i+\frac{1}{2}),~ v_i=0,1,2,\ldots~,
\end{eqnarray}
where $\omega_i$ is the vibrational frequency of the $i$th vibration mode, $i\in\{1,2,3\}$  and $[\omega_i]=$\wn/.

\subsubsection{Rotations}

\water/ has the rotational properties of an asymmetric top with principal moments of inertia $I_a< I_b<I_c$~\cite{Bunker1998}.
The rotational states are indexed by three quantum numbers $J_{K_aK_c}$ where $J=0, 1, 2\ldots$ is the rotational angular momentum quantum number.
$K_a$ and $K_c$ are the absolute values of the projection of $J$ onto the $a$ and $c$ axes, in the limits of a prolate ($I_a=I_b$) and an oblate ($I_b=I_c$) top respectively;  $K_a, K_c\leq J$~\cite{Bunker1998}.
Each rotational state $\ket{J_{K_aK_c}, m}$ is $(2J+1)$-fold degenerate, where $m \in\{ -J,-J+1,\ldots  J\}$ is the projection of $\mathbf{J}$ onto the $z'$ axis of the crystal coordinate frame $C$. 

The energies and the wavefunctions, $\ket{J_{K_aK_c},m}$, of the free rotor Hamiltonian $\mathcal H_{\mathrm {rot}} $ depend on the rotational constants of an asymmetric top, $A_\vv>B_\vv>C_\vv$,~\cite{Bunker1998}
\begin{equation}\label{eq:rot_ham}
\mathcal H_{\mathrm {rot}} = A_\vv \hat J_a^2 + B_\vv \hat J_b^2 + C_\vv \hat J_c^2,
\end{equation}
where $\hat J_i $ are the components of the angular momentum operator $\hat{\mathbf J}^2$ along the principal directions $ a, b$, and $c$. 
The index $\vv$ labels the rotational constants in   vibrational state $\ket{\vv}$.

A molecule-fixed  coordinate system $M$ (axes $x$, $y$ and $z$, Fig.~\ref{fig:MolAxes})  with its origin at the nuclear centre of mass is defined with the following orientations relative to the principal axes of the inertial tensor: $\vect{x}\parallel \vect{b}$,  $\vect{y}\parallel \vect{c}$, and  $\vect{z}\parallel \vect{a}$ where $\vect{y}$ is perpendicular to the H--O--H plane and $\vect{x}$ points towards the oxygen atom.
The rotational wavefunctions   of an asymmetric top in that basis   are \cite{Zare1988,Bunker1998}:
\begin{eqnarray}\label{eq:function_ASymmetric_top_Zare}
\ket{J,k,m,\pm}&=&\left(\ket{J,k,m}\pm\ket{J,-k,m}\right)/\sqrt{2},\\
\ket{J,0,m,\pm} &\equiv&\ket{J,0,m}~\mathrm{ if } \quad k=0,\nonumber
\end{eqnarray}
where $k$ is the projection of $\mathbf{J}$ on the  $a$ axis, axis $z$ of $M$; for $k>0$, $k\in \{ 1, 2, \ldots, J\}$ and each state is doubly degenerate.
Here,  $\ket{J,k,m}$ denotes a normalized rotational  function
\begin{eqnarray}
\ket{J,k,m} & = & \sqrt{\frac{2J+1}{8 \pi^2}} \left[\WignerD{J}{mk}{\Omega_{C\rightarrow M}}\right]^*, 
\end{eqnarray} 
where the Euler angles $\Omega_{C\rightarrow M}$  transform the crystal-fixed coordinate frame $C$  into the molecule-fixed coordinate frame $M$ and $\WignerD{J}{mk}{\Omega}$ is the Wigner rotation matrix element or the Wigner $D$-function~\cite{Varshalovich1988,Zare1988,WignerD}.

The correspondence between the asymmetric top wavefunctions $\ket{J,k,m,\pm}$, Eq.~(\ref{eq:function_ASymmetric_top_Zare}), and  asymmetric top wavefunctions $\ket{J_{K_aK_c}, m}$, is given in \cite{Bunker1998}.
The latter notation of wavefunctions is useful for the symmetry analysis and can be used to relate the wavefunction to   \para/ and \ortho/ states of the water molecule, see   Section~\ref{sec:isomers}.

\subsubsection{Electrostatic interactions}

We assume  two contributions to the electrostatic interaction:
\begin{equation}
\mathcal H_{\mathrm {ES}}=\mathcal{H}_\mathrm{Q}+\mathcal{H}_\mathrm{ed},
\end{equation}
denoting the coupling of the quadrupole and dipole moments of \water/ to the corresponding multipole fields created by the surrounding charges.

{\em Quadrupolar interaction.}
It was shown by Felker \textit{et al.}~\cite{Felker2017} that \Csixty/ molecules in neighbouring lattice sites generate an electric field gradient at the centre of a given \Csixty/ molecule. In \wateratsixty/, the electric field gradient couples to the electric quadrupole moment of the water molecule, lifting the three-fold degeneracy of the $J= 1$ \ortho/-\water/ rotational ground state~\cite{Felker2017}.

The quadrupolar Hamiltonian may be expanded in rank-2 spherical tensors as follows~\cite{Zare1988,Tannoudji2005}:
\begin{eqnarray}\label{eq:QEnergy}
\mathcal{H}_\mathrm{Q} &=& \sum_{m=-2}^{2} (-1)^m V^{(2)}_{-m}Q^{C}_m, 
\end{eqnarray}
where $\{V^{(2)}_{m}\}$ 
are the spherical components of the electric field gradient tensor
and $\{Q^{C}_m\}$ is the quadrupole moment of the water molecule, both expressed in the crystal-fixed coordinate frame $C$.

The experimental value of the \water/ quadrupole moment in the molecule-fixed coordinate frame $M$   is given by 
$\{Q_{xx}, Q_{yy},Q_{zz}\}=\{ -0.13, -2.50, 2.63\}$\,e.s.u$\times$cm$^2$ \cite{Verhoeven1970}.
Since $ |Q_{xx}|<<|Q_{yy}|, |Q_{zz}|$, it holds that  $Q_{zz}\approx -Q_{yy}$, and we may approximate the water quadrupole moment in spherical coordinates as follows:
\begin{eqnarray}\label{eq:Q_diag_water}
\{Q_m^M\} &=&\{\frac{1}{2}( Q_{xx}-Q_{yy}),0, \sqrt{\frac{3}{2}} Q_{zz},0,\frac{1}{2}( Q_{xx}-Q_{yy})\}\nonumber\\
&\approx& Q_{zz} \{\frac{1}{2},0, \sqrt{\frac{3}{2}},0,\frac{1}{2}\}, \quad m=-2,\ldots,+2.\label{eq:water_Q_approx}
\end{eqnarray}

The site symmetry of the \Csixty/ molecule in solid \Csixty/ is $S_6$, with the three-fold symmetry axis  along the cubic $[111]$ axis, which is chosen here to be the $z'$ axis of the crystal coordinate frame $C$.
The spherical tensor component in the frame $C$ is
\begin{equation}\label{eq:QFcomponentsFelkerF2}
\{ V^{(2)C}_m \} = V_\mathrm{Q} \{0,0,1,0,0  \},  
\end{equation}
 where $ m\in\{-2, -1, 0, 1, 2\}$, transforms like the fully symmetric $A_g$ irreducible representation of  the point group $S_6$~\cite{Altmann2011}.
After the transformation of the quadrupole moment (\ref{eq:water_Q_approx}) from the \water/-fixed molecular frame $M$ to the crystal frame $C$ (see Eq.\,\ref{eq:rank_J_transform}), the quadrupolar Hamiltonian (\ref{eq:QEnergy}) is given by: 
\begin{eqnarray}\label{eq:QEnergyWater2}
\mathcal{H}_\mathrm{Q} 	&=& V_\mathrm{Q} Q_{zz}  \left[\sqrt{\frac{3}{2}}  \left[ \WignerD{2}{00}{\Omega_{C\rightarrow M}} \right]^* \right.\\
&&  + \left. \frac{1}{2} \left[ \WignerD{2}{0,-2}{\Omega_{C\rightarrow M}} +\WignerD{2}{02}{\Omega_{C\rightarrow M}} \right]^*\right].\nonumber
\end{eqnarray}

\emph{Dipolar interaction.}
Dielectric measurements of solid \Csixty/ have provided evidence for the existence of electric dipoles in solid \Csixty/~\cite{Alers1992}. 
We assume these electric dipoles can be source of an electric field in the \Csixty/ cage center.

Consider a crystal electric field  $\vect{\mathcal{E}}$ with spherical coordinates $\{\mathcal{E},\phi_E, \theta_E\}$ in the crystal-fixed frame $C$, see  Appendix~\ref{sec:E_static}. For simplicity, we assume a homogeneous crystal field with uniform orientation in the crystal-fixed frame. 
The interaction  of the electric dipole moment with the electric field is given by
\begin{eqnarray}\label{eq:Hed}
\mathcal{H}_\mathrm{ed}&=&-\sum_{m=-1}^1 (-1)^m\mathcal{E}^{E}_{-m} \mu^{E}_m \\
&=&- \sum_{m'=-1}^1   \mathcal{E} \left[ \WignerD{1}{0m'}{\Omega_{E\rightarrow C}} \right]^* \nonumber\\
&&\quad \times \left[\sum_{m^{\prime\prime}=-1}^1  \left[ \WignerD{1}{m'm^{\prime\prime}}{\Omega_{C\rightarrow M}} \right]^*\mu^{M}_{m^{\prime\prime}} \right]\nonumber,
\end{eqnarray}
where the dipole moment in the molecule-fixed frame $M$ is given by
\begin{equation}\label{eq:water_dip_moment_mol_spher}
\{\mu^{M}_m\}=\frac{\mu_x}{\sqrt{2}}\{-1,0, 1\},\quad m\in\{-1, 0, 1\}
\end{equation}
where $\mu_x$ is the permanent dipole moment of water in the Cartesian  coordinates of frame $M$, Fig.~\ref{fig:MolAxes}(a).
Since there are no other anisotropies than  the axially symmetric electric field gradient tensor,   the angle $\phi_E$ is arbitrary and we choose $\phi_E=0$.

The Hamiltonian (\ref{eq:ham_HmVes}) is diagonalized using the basis (\ref{eq:function_ASymmetric_top_Zare})  up to $J\leq 4$ for the ground vibrational state $\ket{000}$ and for the three excited vibrational states $\ket{100}$, $\ket{010}$, and $\ket{001}$.
The  ground state and the three excited vibrational   states are assumed to have  independent rotational constants $A_\vv$, $B_\vv$, and $C_\vv$, where  $\vv=000, 100, 010$ or  $001$.

After separation of coordinates, see Appendix \ref{sec:transition_matrix_element}, the quadrupole and dipole  moments in  equations~(\ref{eq:QEnergyWater2}) and (\ref{eq:Hed}) are replaced by their expectation values,   $\bra{\vv}Q^{M}_{0}\ket{\vv}$,  $\bra{\vv}Q^{M}_{\pm2}\ket{\vv}$ and $\bra{\vv}\mu^{M}_{\pm 1}\ket{\vv}$,  in the ground and in the three excited vibrational states.
We assume for simplicity that the dipole and quadrupole moments of \water/ are independent of the vibrational state $\ket{\vv}$.

\subsubsection{Confined  water translations: spherical oscillator\label{sec:transaltions}}

The translational motion is the center of mass motion and is quantized for a confined molecule. 
The high icosahedral symmetry of the \Csixty/ cavity is close to spherical symmetry and
therefore the translational motion of trapped molecule can be described by the three-dimensional isotropic spherical oscillator model~\cite{Shaffer1944}.
For simplicity we write the  potential in the harmonic approximation~\cite{Tannoudji2005}:
\begin{equation}\label{eq:spher_anharm_potential}
    V(R)=V_2 R^2,
\end{equation}
where $R$ is the displacement of \water/ center of mass from the \Csixty/ cage center.
The \Csixty/ cage is assumed rigid and its center of mass is fixed.

The frequency of the spherical harmonic oscillator is:
\begin{equation}\label{eq:harm_osc_freq}
\omega^0_\mathrm{t}=\sqrt{2V_2/m},     
\end{equation}
where $m$ is the  mass of a  molecule moving in the potential and $[\omega^0_\mathrm{t}]=\mathrm{rad}\,\mathrm{s}^{-1}$.
The energy of spherical harmonic oscillator is quantized,
\begin{equation}\label{eq:Energy_3DHarmonic}
    E_N=\hbar \omega^0_\mathrm{t} (N+\frac{3}{2}),
\end{equation}
where $N$ is the translational quantum number, $N\in \{0, 1, 2, \ldots\}$.
The orbital quantum number $L$ takes values $L=N, N-2,\ldots 1  (0)$  for $N$ odd (even).
Energy  of  the  harmonic spherical  oscillator  does not depend on $L$  and in isotropic approximation there is an additional degeneracy of each $E_N$ level in quantum number $M_L$, taking  $2L+1$ values, $M_L\in \{ -L, -L+1,\ldots L\}$. 

\subsubsection{Nuclear spin isomers: \para/ and \ortho/ water \label{sec:isomers}}

The Pauli principle requires that the total quantum state is antisymmetric with respect to exchange of the two protons in water, which as spin-1/2 particles are  fermions. 
This constraint leads to the existence of two nuclear spin isomers, with total nuclear spin $I = 0$ (\para/-\water/) and $I=1$ (\ortho/-\water/), and different sets of rovibrational states. 
The antisymmetric nature of the quantum state has consequences on the IR spectra: only  \para/ to \para/ and \ortho/ to \ortho/ transitions are allowed.

The \ortho/-\water/ states have odd values of $K_a + K_c$  while the \para/-\water/ states have  even values of $K_a + K_c$ in the ground vibrational state $\ket{000}$, see  Fig.~\ref{fig:MolAxes}(c).
The allowed rotational transitions are depicted in the inset to Fig.~\ref{fig:FIR_spectrum}(a).
The same rule applies to the excited vibrational states $\ket{100}$ and  $\ket{010}$, Fig.~\ref{fig:MolAxes}(c), upper left part. 
However, the rules are inverted for the states $\ket{001}$, $\ket{101}$ and $\ket{011}$, which involve one-quantum excitation of the  asymmetric stretch mode $v_3$. 
In these cases~\cite{Bunker1998}, \para/-\water/ has odd values of  $K_a + K_c$, while \ortho/-\water/ has even values of $K_a + K_c$, Fig.~\ref{fig:MolAxes}(c), upper right part.

The energy difference  between the lowest \para/ rotational state $\ket{0_{00}}$ and the lowest \ortho/ rotational state  $\ket{1_{01}}$ is 2.6\,meV (28\,K)~\cite{Beduz2012}. 
Above  30\,K the ratio of \ortho/ and \para/ molecules is $n_\mathrm{ o}/n_\mathrm{p}\approx 3$.
Hence, if the sample is cooled rapidly to 4\,K, the number of \para/ molecules slowly grows  in the subsequent time interval, while the number of the \ortho/ molecules slowly decreases to the thermal equilibrium  value $n_\mathrm{o}\approx 0$. 
The full conversion takes  several hours \cite{Beduz2012,Mamone2014,Meier2015,Zhukov2020}.

\subsection{Absorption cross-section  of \wateratsixty/\label{sec:abs_cross_section}}
The strengths of the transitions between the rotational states of a polar molecule are  determined by the permanent electric dipole moment of the molecule and by the electric field of the infrared radiation, corrected by the  polarizability of the  medium. 
In principle, the polarizability $\chi$ of the solid depends on the fraction $f$ of \Csixty/ cages which contain a water molecule, $\chi(f)= \chi_{\Csixty/} + f \chi_{\wateratsixty/}$. 
However, we found that 
within the studied range of filling factors, $f=0.1$ to 0.8, the  absorption cross-section of \wateratsixty/ was independent of $f$.
Hence, only the polarizability of solid \Csixty/ is relevant, $\chi\approx  \chi_{\Csixty/}$, 
and the problem  is similar to the optical absorption of an isolated impurity atom in a crystal \cite{Dexter1956}.

Following Ref.~\onlinecite{Dexter1956}, the electric field at the molecule embedded into medium with an index of refraction $\eta$ is  ${\cal E}_\mathrm{eff}={\cal E} (\eta^2 +2 )/3$, where ${\cal E}$ is the electric field of radiation in the vacuum.
The refractive index of solid \Csixty/ is $\eta=2$~\cite{Homes1994}, and hence ${\cal E}_\mathrm{eff}=2{\cal E}$.

In the following discussion, we use the index $k\in\{\mathrm{o},\mathrm{p}\}$ to indicate the \ortho/ or \para/ nuclear spin isomers.
The absorption cross-section~\cite{Bunker1998} for a given nuclear spin isomer $k$, including the effective field correction,  is given by
\begin{eqnarray}\label{eq:cross-section_op}
\sigma_{ji}^{(k)} & = & N_k^{-1}\int_{\mathrm{Line}} \alpha_{ji}^{(k)} (\omega) \mathrm{d} \omega \\
& =& \frac{2\pi^2  }{  h \epsilon_0 c_0 \eta  }\left({\frac{\eta^2 +2  }{ 3  }}\right)^2 \omega_{ji}^{(k)} \left(p_i^{(k)}-p_j^{(k)}\right) S_{ji}^{(k)},\nonumber
\end{eqnarray}
where $c_0$ is the speed of light in vacuum and $\epsilon_0 $ the permittivity of vacuum;
SI units are used  and the frequency $\omega$ is in number of waves per meter, $[\omega]=$m$^{-1}$. 
The integral in (\ref{eq:cross-section_op}) is the area of the absorption line of the transition from the state $\ket{i}$ to $\ket{j}$.

The square of the electric dipole matrix element   is given by 
\begin{equation}\label{eq:dip_moment_aver}
S_{ji}^{(k)} = \frac{1}{3}  \sum^1_{\sigma=-1} |\bra{j} \mu^C_\sigma \ket{i}|^2,
\end{equation}
where  $\ket{i}$ and  $\ket{j}$ are the eigenstates with corresponding energies $E_i^{(k)}$ and $E_j^{(k)}$. 
The symbol $\mu^C_\sigma $ denotes the dipole moment components of a water molecule in the crystal-fixed frame $C$.
This form of $S_{ji}^{(k)}$ is valid for  random  orientation of crystals in the powder sample and does not depend on the polarization of light, see Appendix~\ref{sec:E_radiation}.

The absorption cross-section $	\sigma_{ji}^{(k)}$ was evaluated separately for \para/  and  \ortho/    water.
The concentration of  molecules is $N_k = f N_{\mathrm C_{60}}n_{k}$ , where $n_{k}$ is the \ortho/ (or \para/)   fraction and  $n_{\mathrm p}+n_{\mathrm o}=1$.
The filling factor is denoted by $f$ and the number density of \Csixty/ molecules in solid \Csixty/ is given by $N_{\mathrm C_{60}}=1.419\times 10^{21}$ cm$^{-3}$~\cite{Aoyagi2014}.

$p_i^{(k)}$ and $p_j^{(k)}$ are the probabilities that the initial and final states are thermally populated, 
\begin{equation}\label{eq:Boltzmann_factor}
p_n^{(k)} =\left(Z^{(k)}\right)^{-1} \exp\left(-\frac{E_n^{(k)}-E_0^{(k)}}{k_{\mathrm B}T}\right) ,   
\end{equation}

where the statistical sum  is 
\begin{equation}\label{eq:stat_sum}
Z^{(k)}=\sum_n   \exp\left(-\frac{E_n^{(k)}-E_0^{(k)}}{k_{\mathrm B}T}\right)   
\end{equation}
and $E_0^{(k)}$ is the ground state energy of  \para/ (\ortho/) molecules.
At the low temperatures considered in this work, only the vibrational ground state is significantly populated. 
Thus, the significantly thermally populated states are rotational states in the ground vibrational states, which can be written as linear combinations of basis states in Eq.~(\ref{eq:function_ASymmetric_top_Zare}).
Since we expect that the water molecule is not in a spherically symmetric environment in \wateratsixty/, the degeneracy in quantum number $m$ is lifted, in general.
Therefore, $p_n^{(k)}$ is the thermal population of a non-degenerate rotational state and the eigenstates $\ket{i}$ and $ \ket{j}$ in Eq.~(\ref{eq:dip_moment_aver}) include all possible $m$ values for a given $J$.

When thermal equilibrium is reached between  the \para/ and \ortho/ water,  the fraction of nuclear spin isomer $k$ is
\begin{equation}
n_k(T)=\frac{g^{(k)} Z^{(k)}}{g^{(p)} Z^{(p)}+ g^{(o)} Z^{(o)}}.    
\end{equation}
For spin isomer $k$,  the nuclear spin degeneracy $g^{(k)}=2I+1$, with $I=0$ for \para/  and $I=1$ for \ortho/.

The absorption line areas are calculated from Eq.~(\ref{eq:cross-section_op}), where the matrix elements in Eq.~(\ref{eq:dip_moment_aver}) are between the eigenstates $\ket{\Phi_\mathrm{rot}^{v}}$ of the 
Hamiltonian given by  Eq.~(\ref{eq:ham_HmVes}). 
After separation of coordinates (see Appendix~\ref{sec:transition_matrix_element}), the 
matrix elements in the crystal-fixed coordinate frame $C$ are 
\begin{eqnarray}\label{eq:eldip_matrix_elem_separable_main}
&&\bra{j}\mu^{C}_{\sigma} \ket{i}=\\
&&\quad =\sum_{\sigma'=-1}^{1} \bra{\Phi_\mathrm{rot}^{\vv'}} \WignerD{1}{\sigma\sigma'}{\Omega_{C\rightarrow M}}^* \ket{\Phi_\mathrm{rot}^{000}} \bra{\vv'}     \mu_{\sigma'}^{M}\ket{000},\nonumber
\end{eqnarray}
where the initial state is $\ket{i}=\ket{000}\ket{\Phi_\mathrm{rot}^{000}}$
and \ket{\Phi_\mathrm{rot}^{\vv}} are the linear combinations of states (\ref{eq:function_ASymmetric_top_Zare}).
The dipole moments $\bra{\vv'}     \mu_{\sigma'}^{M}\ket{000}$  are given by  Eq.~(\ref{eq:dip00_M_element}) and  (\ref{eq:dip0i_M_element}) in Cartesian coordinates.

\section{Results and interpretation of spectra \label{sec:results}}
\subsection{Spectra}
The water IR absorption lines were identified unambiguously by taking advantage of the slow \ortho/-\para/ conversion at low temperature.
After rapid cooling from 30\,K to 5\,K, the slow \ortho/-\para/ conversion causes the intensity of the \para/ lines to slowly increase, while the intensity of the \ortho/ lines decreases. 
The water absorption lines are readily identified, and assigned to one of the two spin isomers, by taking the difference between spectra acquired shortly after cooling and spectra acquired after an equilibration time at the lower temperature.

\begin{figure}
	\includegraphics[width=0.48\textwidth]{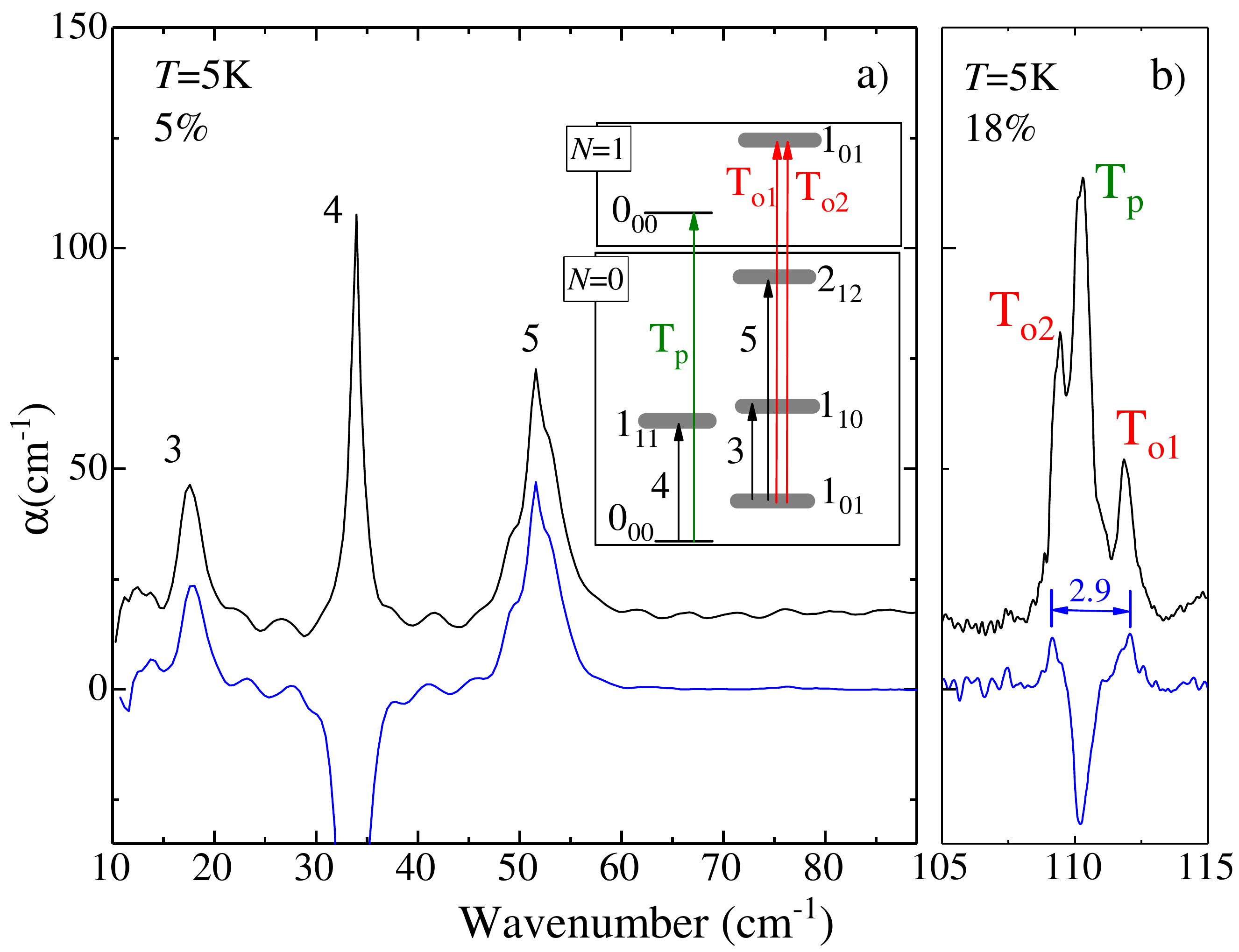}
	\caption{\label{fig:FIR_spectrum}
	    Far-IR absorption spectra of \wateratsixty/ at 5\,K.
		(a) Spectrum $\alpha(0)$ was measured  after the temperature jump  from 30\,K to 5\,K   (black) and the difference  $\Delta \alpha = \alpha(0)-\alpha (\Delta t)$  was measured $\Delta t=44$\,h  later (blue).
		The sample filling factor  $f=0.05$.
		Water rotational transitions corresponding to the absorption lines numbered  3 and 5 (\ortho/-water) and 4 (\para/-water),  are shown in the inset.
		(b)  Spectrum $\alpha(0)$ was measured  after the temperature jump  from 30\,K to 5\,K   (black) and the difference  $\Delta \alpha = \alpha(0)-\alpha (\Delta t)$ was  measured $\Delta t=5$\,h  later (blue).
		The sample filling factor  $f=0.18$.
		 The translational transitions $N=0\rightarrow N=1$ for \para/ (T$_{\mathrm p}$) and  \ortho/ (T$_{\mathrm o1}$, T$_{\mathrm o2}$ ) \wateratsixty/ are shown in the inset to panel (a).
		$N$ is the quantum number of the spherical harmonic oscillator, Eq.\,(\ref{eq:Energy_3DHarmonic}).
	}
\end{figure}

A group of \water/ lines, numbered 3, 4, and 5, is seen below 60\wn/, Fig.~\ref{fig:FIR_spectrum}(a).
These far-IR absorption lines have been reported earlier   and correspond to  the rotational transitions  of \water/ in the \Csixty/ cage~\cite{Beduz2012}.
The rotational  energy levels involved  are shown in the inset to Fig.\,\ref{fig:FIR_spectrum} (a).
Lines 3 and 5 are \ortho/ water rotational transitions starting from the \ortho/ water ground state $\ket{1_{01}}$.
Line 4 is the \para/ water transition from the ground rotational state $\ket{0_{00}}$.
No other rotational transitions were observed at 5\,K which is  consistent with the selection rules for the electric dipole allowed rotational transitions from states $\ket{0_{00}}$ and $\ket{1_{01}}$ \cite{Bunker1998}.

Further \water/ lines are observed around 110\wn/, Fig.~\ref{fig:FIR_spectrum}(b), and in six spectral regions above 600\wn/ as shown  in Figures~\ref{fig:Subl_H2O_1500_5K} to \ref{fig:Subl_H2O_H2OatC60_5000_7000_5K_relaxation}.
Below, we address each wavenumber range separately and assign the spectral lines to the transitions shown in the energy schemes of  Fig.~\ref{fig:MolAxes}(c) and  Fig.~\ref{fig:FIR_spectrum}\,(a).
The absorption lines associated with transitions of free water, labelled 1 to 8, are listed in Table~\ref{tab:IR_transitions}.
Line  assignments are supported by the results of the spectral fitting using the model of a vibrating rotor in a crystal field.

\begingroup
\begin{table*}
\begin{ruledtabular}
\caption{\label{tab:IR_transitions}
The rotational and rovibrational transition frequencies $\omega_{ji}$ and the normalized absorption cross-sections 
$\langle\sigma_{ji}^{(k)}\rangle$, Eq.~(\ref{eq:norm_crosssection}), from the ground \para/, $\ket{0_{00}}$, and ground \ortho/, $\ket{1_{01}}$, rotational state of \wateratsixty/ and of free \water/. 
The initial vibrational state is $\ket{000}$ for all transitions.
The spectral lines are labelled by \#  from 1 to 8, each number associated with the same pair of initial and final rotational states. 
The frequency  $\omega_{ji}$ of lines 3 to 8 in \wateratsixty/ is the intensity-weighted average of line sub-component frequencies.
Gas phase $\omega_{ji}$ are from Ref.~\onlinecite{Tennyson2001}.
}

 \begin{tabular}{ccccrrd{1.10}cd{1.3}c}
	$\ket{v_1v_2v_3}$ & \# &\multicolumn{2}{c}{$\ket{J_{K_aK_c}}$}& \multicolumn{2}{c}{$\omega_{ji}/\wn/$} & \multicolumn{3}{c}{$\langle\sigma_{ji}^{(k)}\rangle/(\sikma/)$}  & \multicolumn{1}{c}{$\langle\sigma_{ji}^{(k)}\rangle(\mathrm{ gas})/\langle\sigma_{ji}^{(k)}\rangle({{\mathrm {C_{60}}}}) $} \\
	\cline{3-4}\cline{5-6}\cline{7-9}
	$j$ & &$i$ &$j$ & $\mathrm{gas}$  &\Csixty/ & \multicolumn{1}{c}{gas} &Ref.& \multicolumn{1}{c}{\Csixty/} &  \\  
	\hline
	000&  3 & $1_{01}$ &   $1_{10}$ & 18.6 & 16.8 & (4.003\pm0.020)10^{-17} & \cite{Rothman1992} & (3.24\pm0.26)10^{-18} & 12.4 \\
	&4& $0_{00}$ &   $1_{11}$ & 37.1 &33.6 & (5.34\pm0.03)10^{-17} &  & (3.02\pm0.19)10^{-18} & 17.4 \\
	&5&  $1_{01}$ &   $2_{12}$ &55.7 & 51.1 & (1.203\pm0.006)10^{-16} & &(7.5\pm0.4)10^{-18} & 16.1 \\
	\hline
	100 &1& $0_{00}$ &  $0_{00}$& {3657.1} &{3573.2}  & & &(3.4 \pm 0.6)10^{-20}   \\
	&2& $1_{01}$ &  $1_{01}$ &&   &   \\
	\cline{2-10}
	
	& 3& $1_{01}$ &  $1_{10}$ & 3674.7 &  3589.1 & (4.39\pm0.17)10^{-19}&\cite{Flaud1975} & (2.89\pm0.03)10^{-19} & 1.52 \\
	& 4 & $0_{00}$ &  $1_{11}$ & 3693.3 &  3606.5 & (2.63\pm0.10)10^{-19}& & (1.93\pm0.14)10^{-19} & 1.35 \\
	& 5 & $1_{01}$ &  $2_{12}$ & 3711.1& 3623.5 & (3.47\pm0.14)10^{-19}& & (1.57\pm0.12)10^{-19} & 2.21 \\
	\hline
	010 &1&  $0_{00}$ &  $0_{00}$& {1594.8} & {1569.3}  & & & (2.3\pm0.6)10^{-19}   \\
	&2& $1_{01}$ &  $1_{01}$ &&   &   \\
	\cline{2-10}
	
	&3& $1_{01}$ &  $1_{10}$ & 1616.7 &  1588.5 & (1.51\pm0.03)10^{-17}  &\cite{Toth1991} & (1.03\pm0.11)10^{-18} & 14.7 \\
	&4& $0_{00}$ &  $1_{11}$ & 1635.0 &  1605.3 & (1.045\pm0.021)10^{-17}  & & (6.79\pm0.23)10^{-19} & 15.4 \\
	&5& $1_{01}$ &  $2_{12}$ & 1653.3 & 1623.4 &  (1.57\pm0.06)10^{-17} & & (1.26\pm0.09)10^{-18} & 12.4 \\
	\hline
	
	001&6& {$0_{00}$} &  $1_{01}$ & 3779.5 & 3682.1 &(7.18\pm0.29)10^{-18}& \cite{Flaud1975} &(1.73\pm0.21) 10^{-18} &4.14 \\
	&7& $1_{01}$ &$0_{00}$ &  3732.1 & 3637.4 &  (7.6\pm0.3)10^{-18} && (1.31\pm0.12) 10^{-18}& 5.77\\
	&8& $1_{01}$ & $2_{02}$ & 3801.4&  3703.7 & (1.37\pm0.05)10^{-17}& & (2.18\pm0.10) 10^{-18} & 6.27\\ 
	
	\hline
	011&6&  $0_{00}$ &  $1_{01}$ &5354.9 & 5228.0 &(8.68\pm0.08)10^{-19}& \cite{Toth2005} & (1.15\pm0.06) 10^{-19} & 7.57 \\
	&7& $1_{01}$ &$0_{00}$ & 5307.5 &  5183.6 &  (8.93\pm0.09)10^{-19} && (8.22\pm0.04) 10^{-20} & 10.9\\
	
	&8& $1_{01}$ & $2_{02}$ & 5376.9 & 5249.2 & (1.71\pm0.03)10^{-18}& & (1.34\pm0.06)10^{-19}  &12.7\\
	\hline
	101&  6 &$0_{00}$ &  $1_{01}$ & 7273.0 &  7088.4 & (5.19\pm0.16)10^{-19}& \cite{Toth1994} & (1.261\pm0.021) 10^{-20} & 41.1 \\ 
	&7& $1_{01}$ & $0_{00}$ & 7226.0 &  7043.5 &  (5.39\pm0.22)10^{-19} &&  (1.044\pm0.012) 10^{-20}& 51.6\\
	&8& $1_{01}$ & $2_{02}$ & 7294.1 & 7109.3 & (1.00\pm0.04)10^{-18} && (1.85\pm0.05) 10^{-20} & 54.1\\
	\hline
	020&4& {$0_{00}$} &  $1_{11}$& 3196.1& 3142.0 &  (6.61\pm0.26)10^{-20}& \cite{Toth1973} &  (5.5\pm0.4) 10^{-21} & 12.0
	
	\\
	
\end{tabular}

\end{ruledtabular}
\end{table*}
\endgroup

\subsubsection{Translational transitions}

A group of absorption lines around 110\wn/ is shown in Fig.~\ref{fig:FIR_spectrum}\,(b). These lines do not correspond to any known  water rotational transitions.
We assign these peaks to the translational transitions ($N=0\rightarrow N=1$) of  \para/- and  \ortho/-\water/, corresponding to the quantized centre-of-mass vibrational motions of the water molecules in the encapsulating \Csixty/ cages. Here $N$ denotes the quantum number of a spherical harmonic oscillator \cite{Tannoudji2005}.

The assignment of these peaks to water centre-of-mass translational oscillations is supported by the presence of  lines  at  1680\wn/, visible in the difference spectrum shown in the right-hand inset to Fig.~\ref{fig:Subl_H2O_1500_5K}.  
These lines are 110\wn/  higher than  the vibrational transitions 1 and 2 of  the  ${v_2}$ mode and correspond to the simultaneous excitation of the $v_2$ vibration and the translational modes. 
A similar combination has been observed in \htwoatsixty/ where a group of lines between 4240 and 4270\wn/ is the translational sideband to \Htwo/ stretching vibration~\cite{Mamone2009}.

The translational side peak  of the $v_1$ vibrational mode is expected at about $3683$\wn/.
However, this frequency coincides with a strong rovibrational absorption line 6 of the $v_3$ mode (Fig.~\ref{fig:Subl_H2O_3500_5K}), which probably obscures the 3683\wn/ translational side peak of the $v_1$ vibration.

The translational \ortho/ transitions display a splitting of 2.9\wn/  in the ground vibrational state, see Fig.\ref{fig:FIR_spectrum}(b),  and 2.7 \wn/  in the excited vibrational state $\ket{010}$, see Fig.~\ref{fig:Subl_H2O_3500_5K}.
These splittings may be attributed to the coupling between the water translation and rotation, associated with the interaction of the non-spherical rotating water molecule with the interior of the \Csixty/ cage. 
Spectral structure of this type has been analysed in detail for the case of \Htwoatsixty/~\cite{Mamone2009,Ge2011, Ge2011D2HD}.
The simplified theoretical model used here does not include translation-rotation coupling and cannot explain these splittings. 
A theoretical analysis of the translational peaks will be given in a later paper.

\subsubsection{ Vibrational and rovibrational transitions}

	\begin{figure}
	\includegraphics[width=0.47\textwidth]{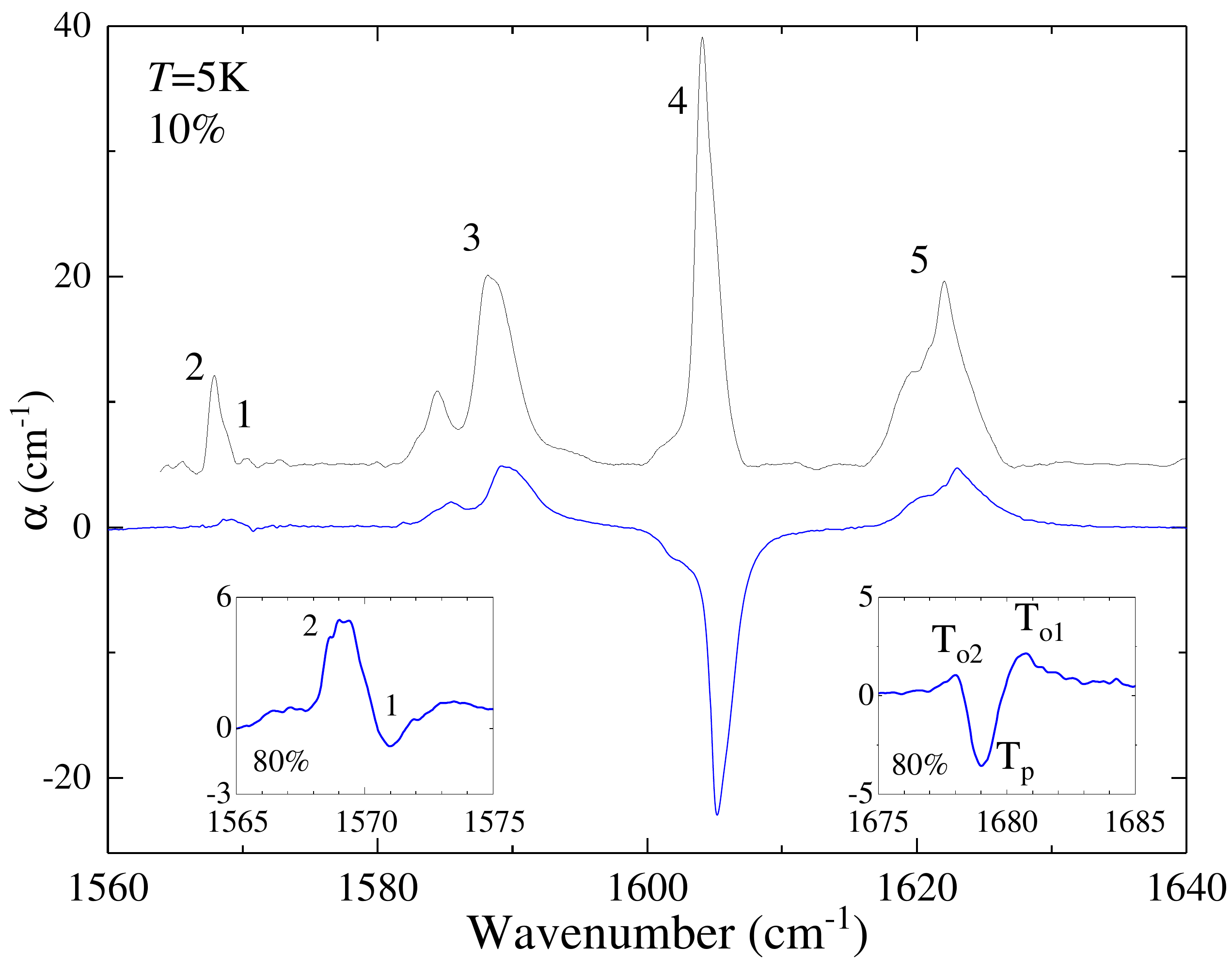}
	\caption{\label{fig:Subl_H2O_1500_5K} 
		Absorption spectra of \wateratsixty/ vibrational, rovibrational and vibration-translational transitions of bending vibration  $v_2$ at 5\,K. 
		Spectrum $\alpha(0)$ was measured  after the temperature jump  from 30\,K to 5\,K   (black) and the difference  $\Delta \alpha = \alpha(0)-\alpha (\Delta t)$ was  measured $\Delta t=3$\,h  later (blue).
		 Sample  filling factor $f=0.1$.
		 Lines numbered 1 and 2 are pure vibrational transitions  and  3, 4, and 5 are rovibrational transitions, Fig.~\ref{fig:MolAxes} (c).
		Left inset show the \para/, line 1, and \ortho/,  line 2,  components of the pure vibrational transitions difference spectrum  with time delay  1\,h  of the  $f=0.8$ sample.
		The    differential spectrum of ${v_2}$  vibration-translational transitions $\mathrm {T_{p}}$, $\mathrm {T_{o1}}$ and $\mathrm {T_{o2}}$, at 110\wn/ higher frequency from 1 and 2,  is  in the right inset.
		}
\end{figure}

	\begin{figure}
	\includegraphics[width=0.48\textwidth]{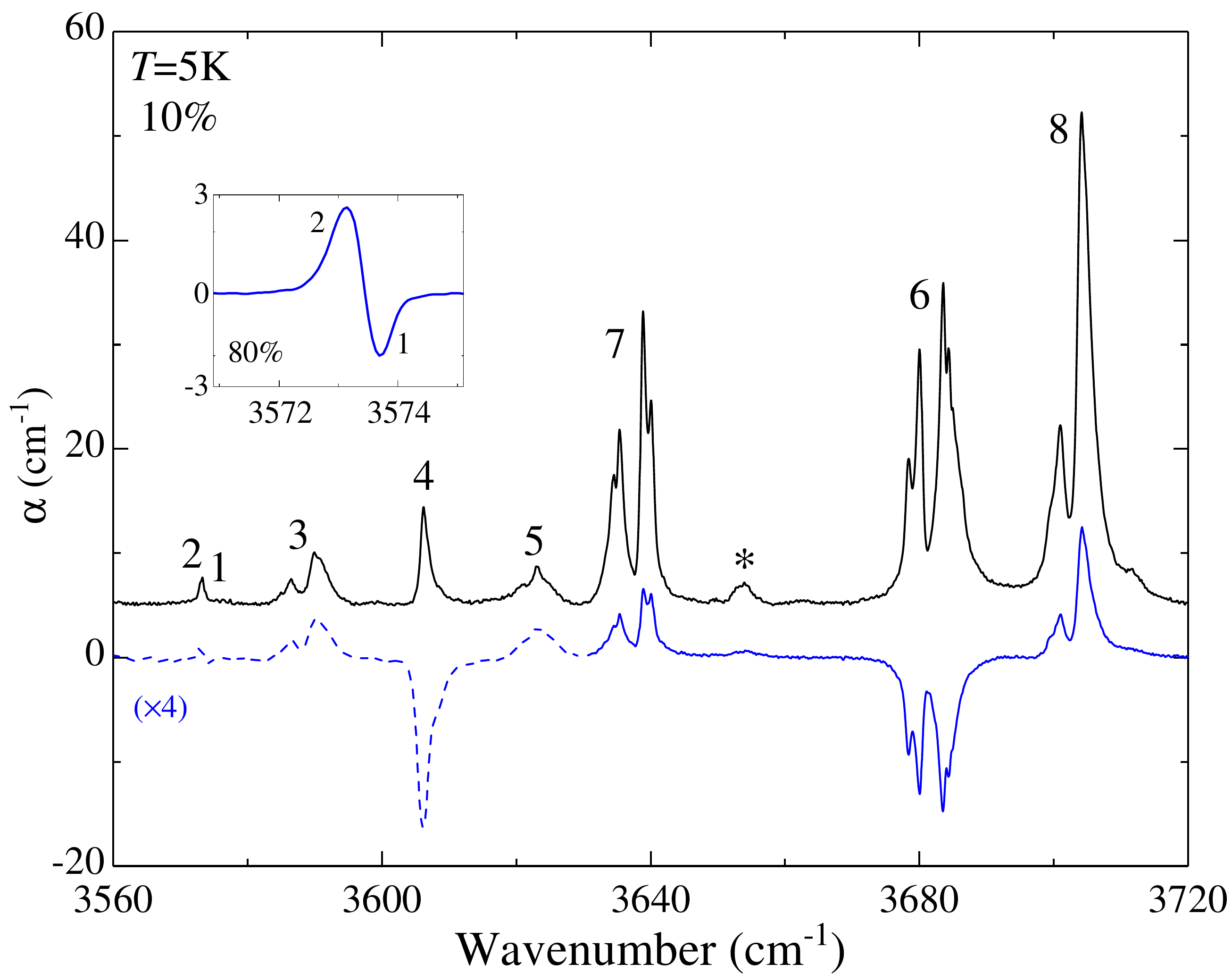}
	\caption{\label{fig:Subl_H2O_3500_5K} 
		Absorption spectra of \wateratsixty/ vibrational, rovibrational and vibration-translational transitions of symmetric stretching, $v_1$, and anti-symmetric stretching vibration  $v_3$ at 5\,K.
		Spectrum $\alpha(0)$ was measured  after the temperature jump  from 30\,K to 5\,K   (black) and the difference  $\Delta \alpha = \alpha(0)-\alpha (\Delta t)$ was  measured $\Delta t=3$\,h  later (blue dashed line multiplied by 4 and blue line).
		 Sample  filling factor $f=0.1$
		 Lines numbered 1 and 2 are pure vibrational transitions  and  3, 4, and 5 are rovibrational transitions of mode ${v_1}$ and lines 6, 7, 8 are rovibrational transitions of $v_3$,  Fig.~\ref{fig:MolAxes} (c).
		 (*)  marks the $v_3$ rovibrational transition at 3654\wn/ from the thermally excited rotational  state $\ket{000}\ket{1_{10}}$ to rovibrational state $\ket{001}\ket{1_{11}}$.
		Inset shows the \para/, line 1, and the \ortho/,  line 2, components of the pure vibrational transition of mode ${v_1}$ of the   $f=0.8$ sample as the difference spectrum with time delay 1\,h.
		}
\end{figure}

\begin{figure}
	\includegraphics[width=0.48\textwidth]{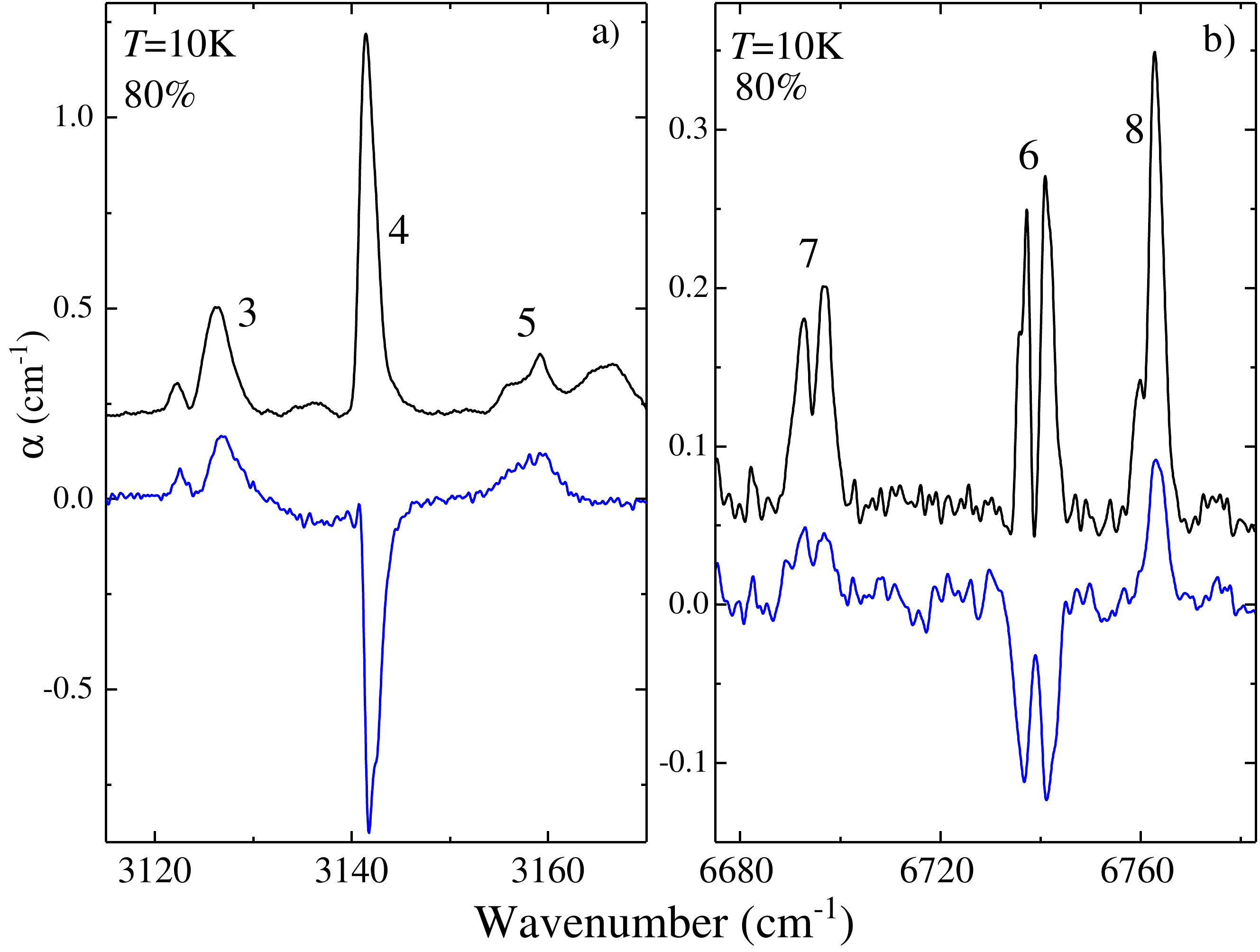}
	\caption{\label{fig:2v2_10K} 
	Absorption spectra of \wateratsixty/ at 10\,K.  
	(a) Overtone,   $2\omega_{2}$, and (b) combination, $2\omega_{2}+\omega_{3}$,   rovibrational transitions.
	Spectrum $\alpha(0)$ was measured  after the temperature jump  from 45\,K to 10\,K   (black) and the difference  $\Delta \alpha = \alpha(0)-\alpha (\Delta t)$  was measured $\Delta t=2.75$\,h  later (blue).
		The numbers 3 -- 8 label  the transitions shown in   Fig.~\ref{fig:MolAxes}\,(c).
		Sample filling factor $f=0.8$.
		} 
\end{figure}

\begin{figure}
	\includegraphics[width=0.48\textwidth]{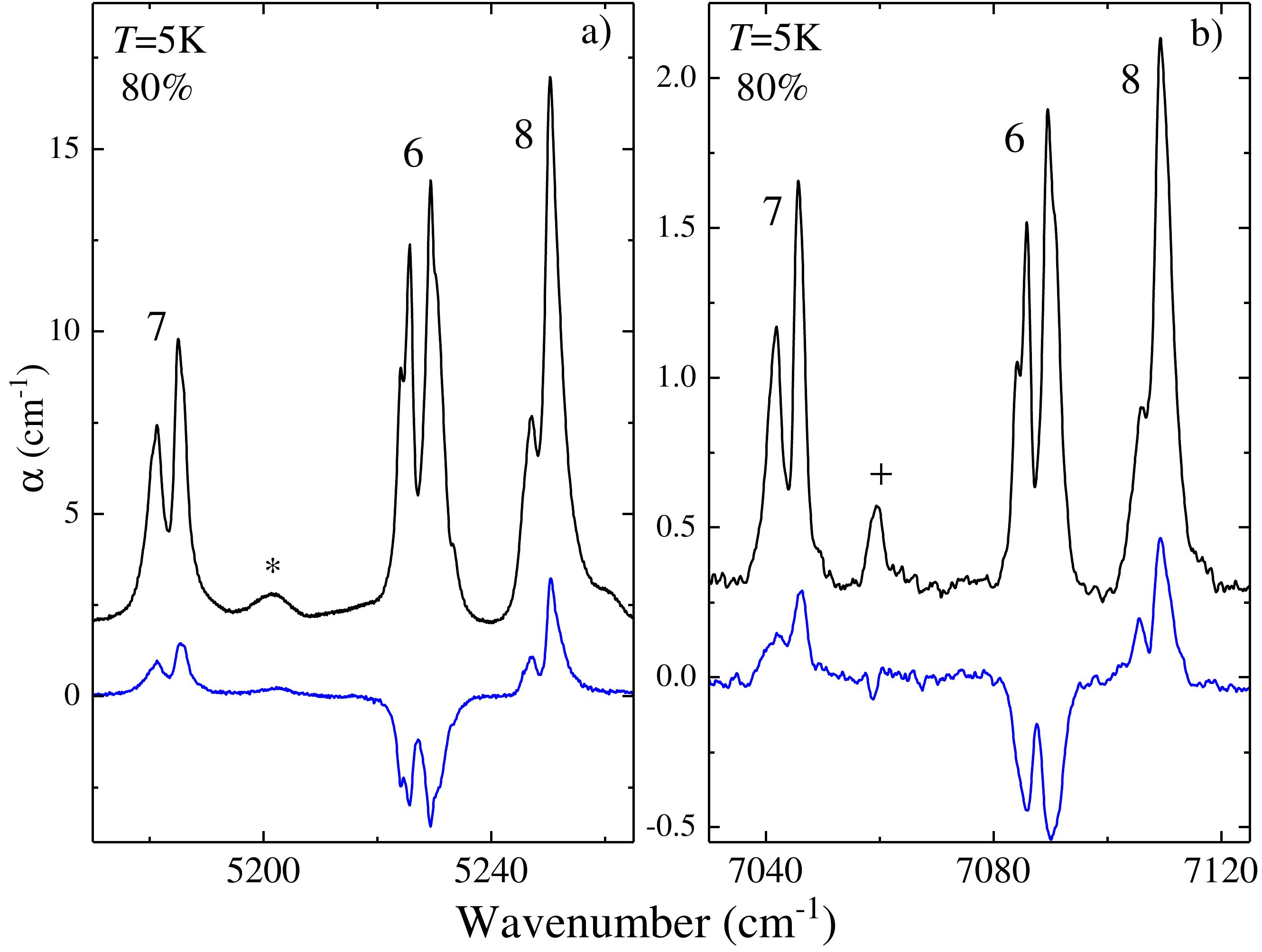}
	\caption{\label{fig:Subl_H2O_H2OatC60_5000_7000_5K_relaxation} 
		Absorption spectra of \wateratsixty/ at 5\,K.  
		(a) Rovibrational combination  $\omega_2+\omega_3$  and (b)  $\omega_1+\omega_3$  transitions.
			Spectrum $\alpha(0)$  was measured  after the temperature jump  from 30\,K to 5\,K   (black) and the difference  $\Delta \alpha = \alpha(0)-\alpha (\Delta t)$  was measured $\Delta t=0.83$\,h  later (blue).
		The numbers 6, 7,  8 label  the transitions shown in   Fig.~\ref{fig:MolAxes}\,(c).
		(*) in  (a) at 5202\wn/ marks the $\omega_2+\omega_3$ rovibrational transition  from the thermally excited rovibrational  state $\ket{000}\ket{1_{10}}$ to $\ket{011}\ket{1_{11}}$.
		The $2\omega_1$ rovibrational transition at 7059\wn/, marked  (+) in (b), is from  $\ket{000}\ket{0_{00}}$ to $\ket{200}\ket{1_{11}}$.
		Sample filling factor $f=0.8$.
		 } 
\end{figure}

\begin{table}
	\begin{ruledtabular}
		\caption{\label{tab:vib_frequencies}
			Frequencies of symmetric stretching ($\omega_1$), symmetric bending ($\omega_2$), and asymmetric stretching ($\omega_3$) modes and of their combinations $2\omega_2$, $\omega_2+\omega_3$, $2\omega_2+\omega_3$, $2\omega_1$, and $\omega_1+\omega_3$ measured from the ground vibrational state $\ket{000}$ for  \wateratsixty/ (this work) and for free \water/~\cite{Tennyson2001}.
			\ortho/ and \para/ components of pure vibrational transitions $v_1$ and $v_2$  are indicated by superscripyts o and p.
			The  frequencies of transitions involving  $v_3$  in \wateratsixty/ are the average of  frequencies of transition 6 and 7, see Fig.~\ref{fig:Subl_H2O_3500_5K}, \ref{fig:2v2_10K}\,(b) and \ref{fig:Subl_H2O_H2OatC60_5000_7000_5K_relaxation}.
			The overtone $2\omega_2$  is estimated from the frequency of line 4, Fig.~\ref{fig:2v2_10K}(a), $2\omega_2+E_{1_{11}}=3141$\wn/, where $E_{1_{11}}=A+C$ is the  energy of rotational state $\ket{1_{11}}$~\cite{Zare1988}, and $A=A_{010}$, $C=C_{010}$ from Table~\ref{tab:parameters}.
			The $2\omega_1$ overtone frequency is estimated from the frequency of \para/ line $2\omega_1+E_{1_{11}}=7059$\wn/, Fig.~\ref{fig:Subl_H2O_H2OatC60_5000_7000_5K_relaxation}(b), where $E_{1_{11}}=A_{100}+C_{100}$.
			Shift $\Delta\omega= \omega_{\mathrm{C}_{60}}-\omega_{\mathrm{gas}}$. 
		} 
		\begin{tabular}{l|d{4.2}d{4.2}d{4.2}d{3.4}}
			$\ket{v_1v_2v_3}$ & \multicolumn{1}{c}{\wateratsixty/} &
			\multicolumn{1}{c}{$\mathrm{free}$\,  \water/} &
			\Delta\omega &
			\multicolumn{1}{c}{$\Delta\omega/\omega_{\mathrm{gas}}$}  \\ 
			& \multicolumn{1}{c}{\wn/} & \multicolumn{1}{c}{\wn/}& \multicolumn{1}{c}{\wn/}&  \\\hline
			$\ket{010}$ & 1569.2^\mathrm{o} & 1594.8 & -25.6& -0.016\\
			& 1571.0^\mathrm{p} & &&\\
			$\ket{100}$ & 3573.2^\mathrm{o} & 3657.1 & -83.9 & -0.023\\
			& 3573.7^\mathrm{p} && & \\
			$\ket{001}$ & 3659.6 & 3755.3 & -95.7 & -0.025 \\
			$\ket{020}$ &3105.5 & 3151.6 &-46.1 & -0.015\\
			$\ket{011}$ & 5205.2 & 5331.3 & -126.1& -0.024  \\
			$\ket{021}$&6716.6&6871.5&-154.9&-0.023 \\
			$\ket{200}$&7027.0&7201.5&-173.5&-0.024 \\
			$\ket{101}$ & 7065.8  & 7249.8 & -184.0& -0.025\\	
		\end{tabular}
	\end{ruledtabular}
\end{table}

The vibrational and rovibrational transitions are shown in Fig.~\ref{fig:Subl_H2O_1500_5K}.
The three  major features that distinguish the spectrum of \wateratsixty/ from the spectrum of free water are as follows:
	
\emph{1. Pure vibrational transitions.}  Absorptions corresponding to pure vibrational transitions, i.e. without simultaneous rotational excitation, are present around ${\omega_1}=3573$ \wn/ and ${\omega_2}=1570$\wn/.
Both features are split into two components, labeled 1 and 2,  identified from the difference spectra,  Fig.~\ref{fig:Subl_H2O_1500_5K} and \ref{fig:Subl_H2O_3500_5K}, as \para/ (1) and \ortho/ (2) transitions. 

The transition 1 is a transition from the ground vibrational state to the excited vibrational state without a change of rotational state  $\ket{0_{00}}$. Transition 2 is a vibrational excitation without change in the rotational state $\ket{1_{01}}$. The corresponding transitions are forbidden for an isolated water molecule since the corresponding matrix element is zero~\cite{Bunker1998}. 
As discussed below, the presence of pure vibrational transitions 1 and 2 is consistent with the presence of an electric field in solid \wateratsixty/.

The \ortho/-\para/ splitting, i.e. the separation of lines 1 and 2, is  0.5\wn/ for the  $v_1$ vibrational mode, and  1.8\wn/ for the $v_2$ vibrational mode, 

\emph{2. Spectral splittings.} 
The rovibrational transitions 3 to 8 are split into two or more components (see Figs.~\ref{fig:Subl_H2O_1500_5K} and \ref{fig:Subl_H2O_3500_5K}).
These splittings, as in the case of the rotational transitions, are absent for a water molecule  in the gas phase.
Moreover, transitions  3, 4, and 5 have the same splitting pattern as the rotational transitions in the ground vibrational state, also labeled  3, 4, and 5, see Fig.~\ref{fig:FIR_spectrum}.
Transitions  6 and 7 are special since they are between the $\ket{0_{00}}$ and  $\ket{1_{01}}$ rotational states and thus reflect directly the splitting of the triply degenerate $\ket{1_{01}}$ state either in the ground vibrational state, transition 7, or in the excited vibrational state, transition 6,  Fig.~\ref{fig:Subl_H2O_3500_5K}. 

We assign a  weak \ortho/ line at 3654\wn/, marked by * in Fig.~\ref{fig:Subl_H2O_3500_5K}, to  the $v_3$ rovibrational transition from the thermally excited rovibrational state $\ket{000}\ket{1_{10}}$ to $\ket{001}\ket{1_{11}}$.
This assignment is further confirmed by calculating the transition frequency \cite{Zare1988} with the parameters from Table~\ref{tab:parameters}: 
$\omega_3 + E_{1_{11}} - E_{1_{10}} = \omega_3 +A_{001}+ C_{001} -(A_{000}+B_{000})=3655$\wn/.

\emph{3. Red shifts.} 
The  frequencies of  vibrations are  red-shifted relative to   free \water/. 
The stretching  mode frequencies are red-shifted by about 2.4\%,  while the  bending mode  frequencies are red-shifted by about   1.6\%,  see  Table~\ref{tab:vib_frequencies}.

\subsubsection{Overtone and combination rovibrational transitions}

Overtone  and  combination vibrational transitions where  two   vibrational quanta are excited  are presented for the $2\omega_{2}$ transition in Fig.~\ref{fig:2v2_10K}\,(a) and for the $\omega_{2} +\omega_{3}$ and $\omega_{1} +\omega_{3}$ transitions in  Fig.~\ref{fig:Subl_H2O_H2OatC60_5000_7000_5K_relaxation}\,(a) and (b). 
A three-quantum transition,  $2\omega_{2} +\omega_{3}$, is shown in Fig.~\ref{fig:2v2_10K}\,(b). 
Rotational levels involved are sketched  in Fig.~\ref{fig:MolAxes}\,(c).
Again,  the splitting pattern of each higher order rovibrational transition is similar to the splitting of  rotational transition with $\Delta v_i=0$ (Fig.~\ref{fig:FIR_spectrum}) and  rovibrational transitions with $\Delta v_i=+1$ (Fig.~\ref{fig:Subl_H2O_1500_5K} and \ref{fig:Subl_H2O_3500_5K}).

We assign a line marked by ``$+$"  at 7059 \wn/  to   2$\omega_{1}$   plus \para/-\water/ rotational transition, $0_{00} \rightarrow 1_{11}$,  Fig.~\ref{fig:Subl_H2O_H2OatC60_5000_7000_5K_relaxation}(b).
Another two rotational side peaks of $2\omega_1$ are \ortho/ transitions 3 and 5 expected at $2\omega_1+E_{1_{10}}-E_{1_{01}}=7042$\wn/  and at $2\omega_1+E_{2_{12}}-E_{1_{01}}=7076$\wn/,
where $E_{1_{10}}-E_{1_{01}}= A-C$ and $E_{2_{12}}-E_{1_{01}}= A+3C$~\cite{Zare1988}, with the approximation $A=A_{200}\approx A_{100}$, $C=C_{200}\approx C_{100}$. 
The numerical values of $A_{100}$ and  $C_{100}$ are taken from Table~\ref{tab:parameters}.
The first line overlaps with line 7, rovibrational transition of $\omega_1 + \omega_3$. 
The second line  is not observed but this could be due to the low intensity of \ortho/ line 5 relative to  the \para/ line 4, see for example  Fig.~\ref{fig:2v2_10K}\,(a).

All two- and three-quantum vibrational transitions are red-shifted approximately  by 2.4\% except  the $2\omega_2$  where the red shift is 1.5\%, see Table~\ref{tab:vib_frequencies}.

\subsection {Spectral fitting  with a quantum mechanical model}\label{sec:SpectralFitting}

A  synthetic spectrum consisting of Gaussian lines with full width at half maximum  1.5\wn/ was calculated from the experimental normalized line areas $\langle A_{ji}^{(k)}\rangle$, Eq.\,(\ref{eq:line_area_norm}),  using $f=1$, $n_\mathrm{o}=0.7$, $n_\mathrm{p}=0.3$ and $T=5$\,K.
The parameters of the model Hamiltonian (from Section \ref{sec:quantum_model}) and the transition dipole moments (from Section \ref{sec:abs_cross_section}) were determined with a non-linear least squares method by minimizing the difference of synthetic and  modeled spectrum  squared, Appendix~\ref{sec:fittingANDerror}.
The  reported parameter error is the average of errors calculated with $+\delta a_\nu$ and $-\delta a_\nu$ in Eq.~(\ref{eq:V_matrix_contiinum}), where the parameter variation of the $\nu$-th parameter  at its best value $\parvec/_{\mathrm{min}}^\nu$ is $|\delta a_\nu | = 0.005 \,\, \parvec/_{\mathrm{min}}^\nu$.
The fit was applied to the rotational transitions in the ground vibrational state $\ket{000}$  and to the rovibrational transitions from the ground state  to the vibrational states $\ket{010}$, $\ket{100}$,  and $\ket{001}$.
In total, $45$ absorption lines were fitted.

\begin{table}
	\begin{ruledtabular}
			\caption{\label{tab:parameters}
			Parameters obtained from the quantum mechanical model fit of IR spectra of \wateratsixty/ at $T=5$\,K, Fig.~\ref{fig:Simulations}.
			The vibration frequencies $\omega_i$, rotational constants $A_\vv$, $B_\vv$, $C_\vv$, and quadrupolar energy $V_\mathrm{Q} Q_{zz}$ are in units of \wn/; 
			electric field $\mathcal{E}$ in $10^6$Vm$^{-1}$, $\theta_E$ in radians and the dipole moment $\mu^x$ (\ref{eq:dip00_M_element}) and the transition dipole moments   $\mu^{x,z}_{0i}$ (\ref{eq:dip0i_M_element})	in D.
			It is assumed that  $\mu^x$ and $Q_{zz}$ do not depend on the vibrational state.
			The parameters with zero error were not fitted.
		} 
\begin{tabular}{l|d{3.10}d{3.10}}
		Parameter & \multicolumn{1}{c}{Value} & \multicolumn{1}{c}{Error}\\ \hline

$n_{\mathrm o}$  &  0.7 &  0\\

$f$  &  1 &  0\\

$ \omega_1$  &  3574.1 &  0.3\\

$ \omega_2$  &  1569.2 &  0.3\\

$ \omega_3$  &  3659.9 &  0.9\\

$A_{000}$  &  24.15 &  0.17\\

$B_{000}$  &  15.3 &  0.8\\

$C_{000}$  &  8.48 &  0.07\\

$A_{100}$  &  23.1 &  0.3\\

$B_{100}$  &  14.3 &  0.8\\

$C_{100}$  &  8.50 &  0.09\\

$A_{010}$  &  26.7 &  0.3\\

$B_{010}$  &  14.6 &  0.9\\

$C_{010}$  &  8.81 &  0.07\\

$A_{001}$  &  26 &  6\\

$B_{001}$  &  15 &  3\\

$C_{001}$  &  8.2 &  1.8\\

$\mu^x$  &  0.474 &  0.008\\

$\mu^x_{01}$  &  1.031\times 10^{-2} &  0.021\times 10^{-2} \\

$\mu^x_{02}$  &  3.40\times 10^{-2} &  0.05\times 10^{-2}\\

$\mu^z_{03}$  &  2.83 \times 10^{-2} &  0.07\times 10^{-2}\\

$\mathcal{E}$  &  110 &  5\\

$\theta_E$  &  1.4 &  1.0\\

$V_\mathrm{Q} Q_{zz}$  &  -5.0 &  0.5\\
		\end{tabular}

	\end{ruledtabular}
\end{table}

\begin{figure*}
\includegraphics[width=0.45\textwidth]{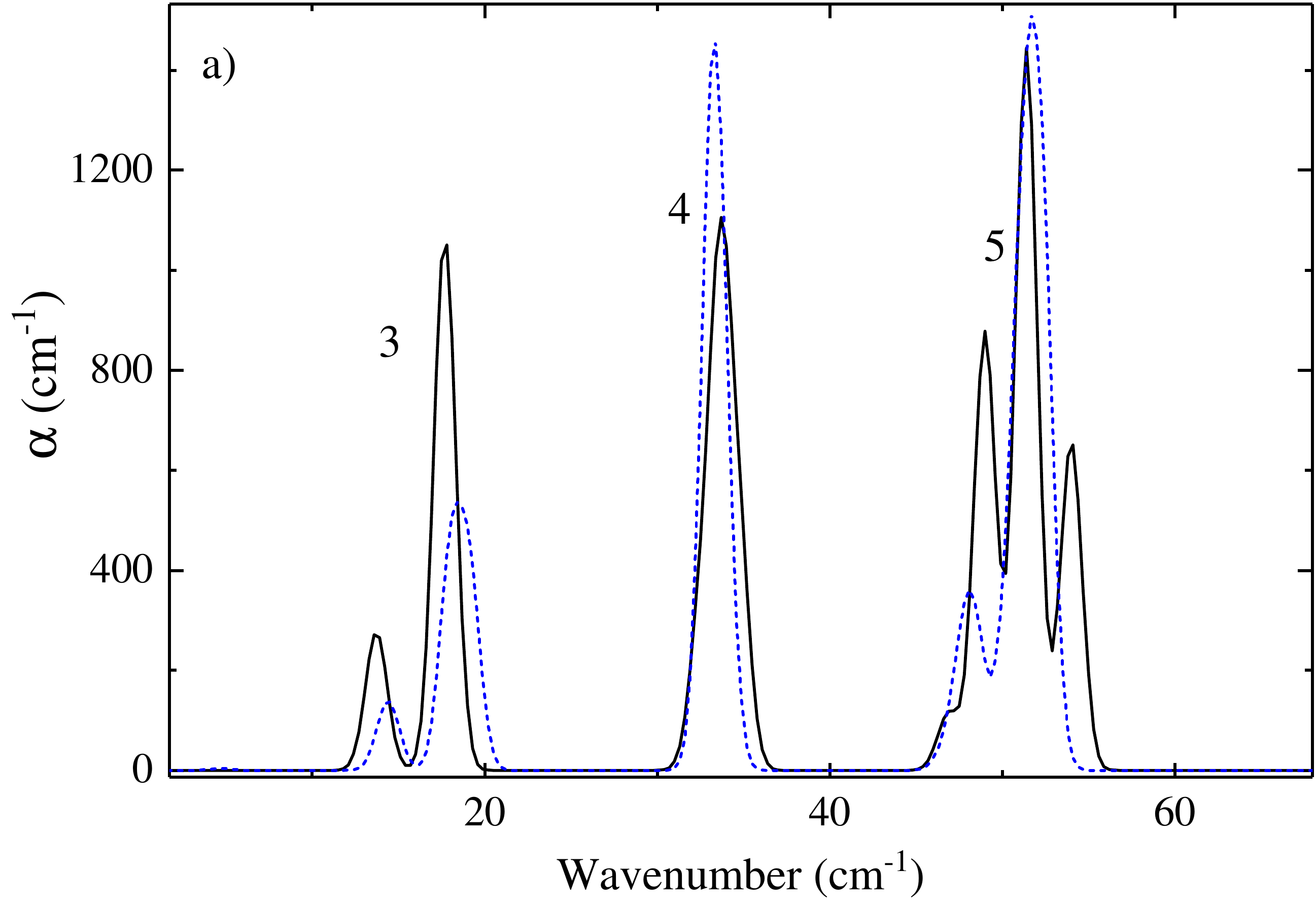}
\includegraphics[width=0.45\textwidth]{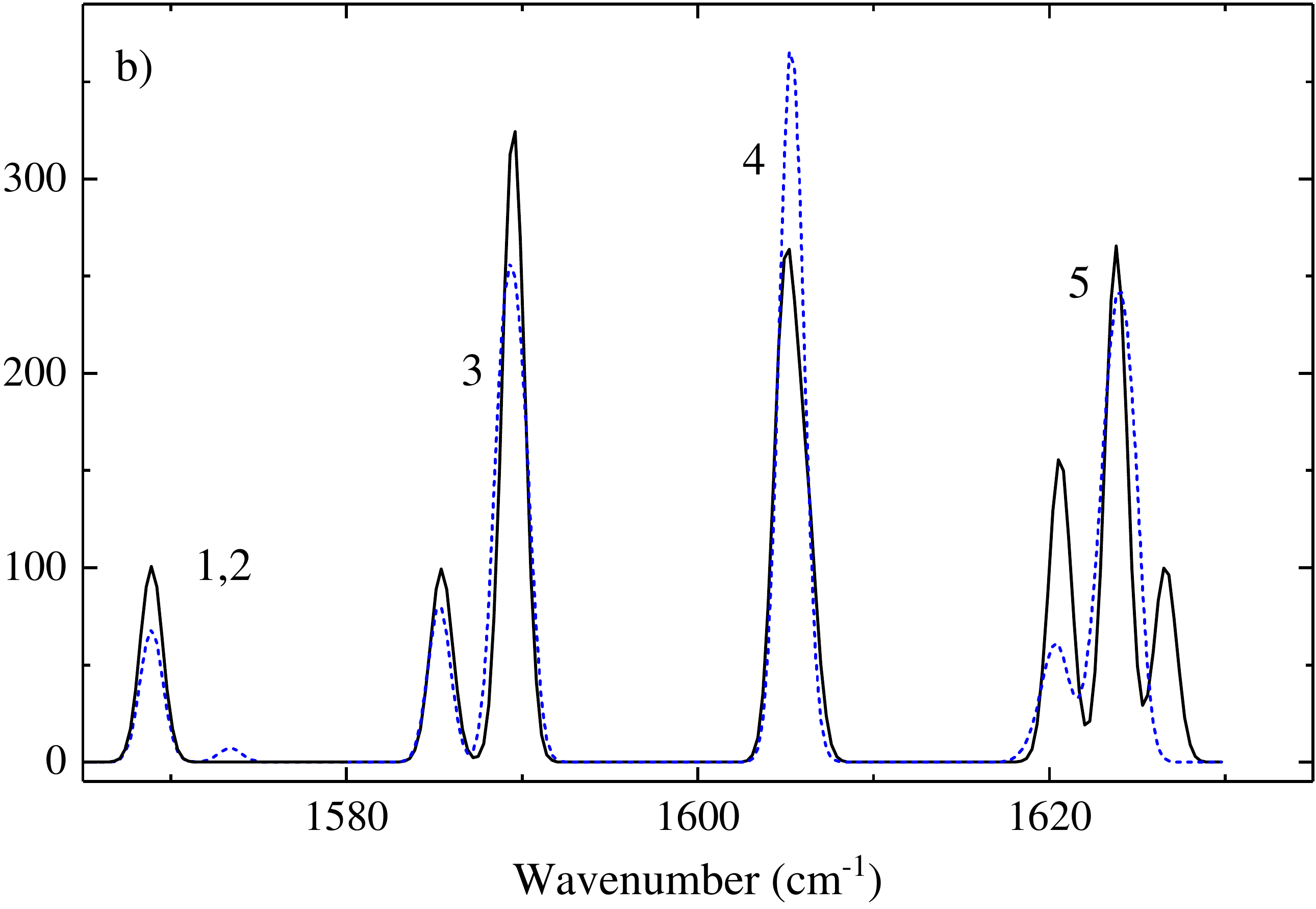}
\includegraphics[width=0.45\textwidth]{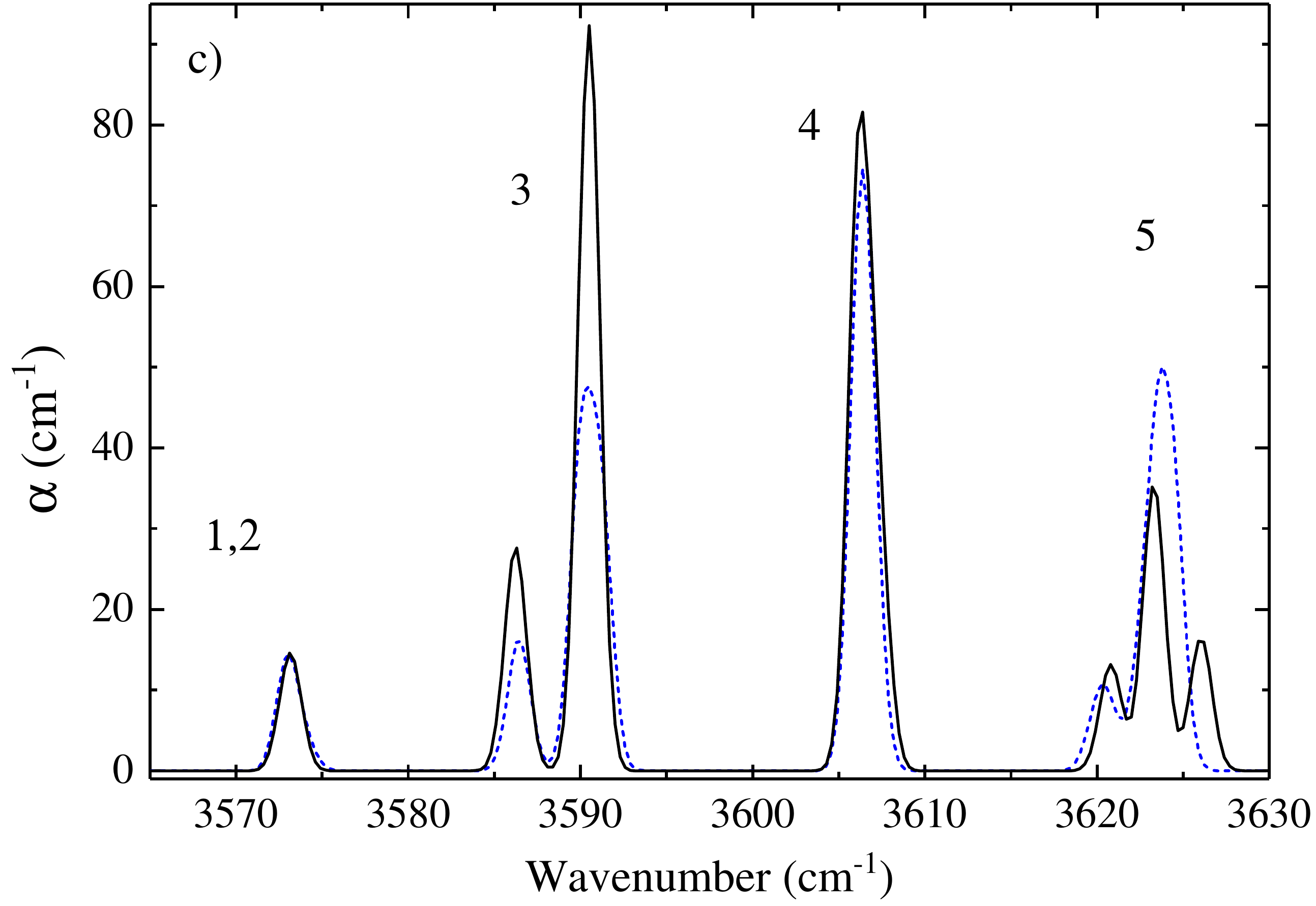}
\includegraphics[width=0.45\textwidth]{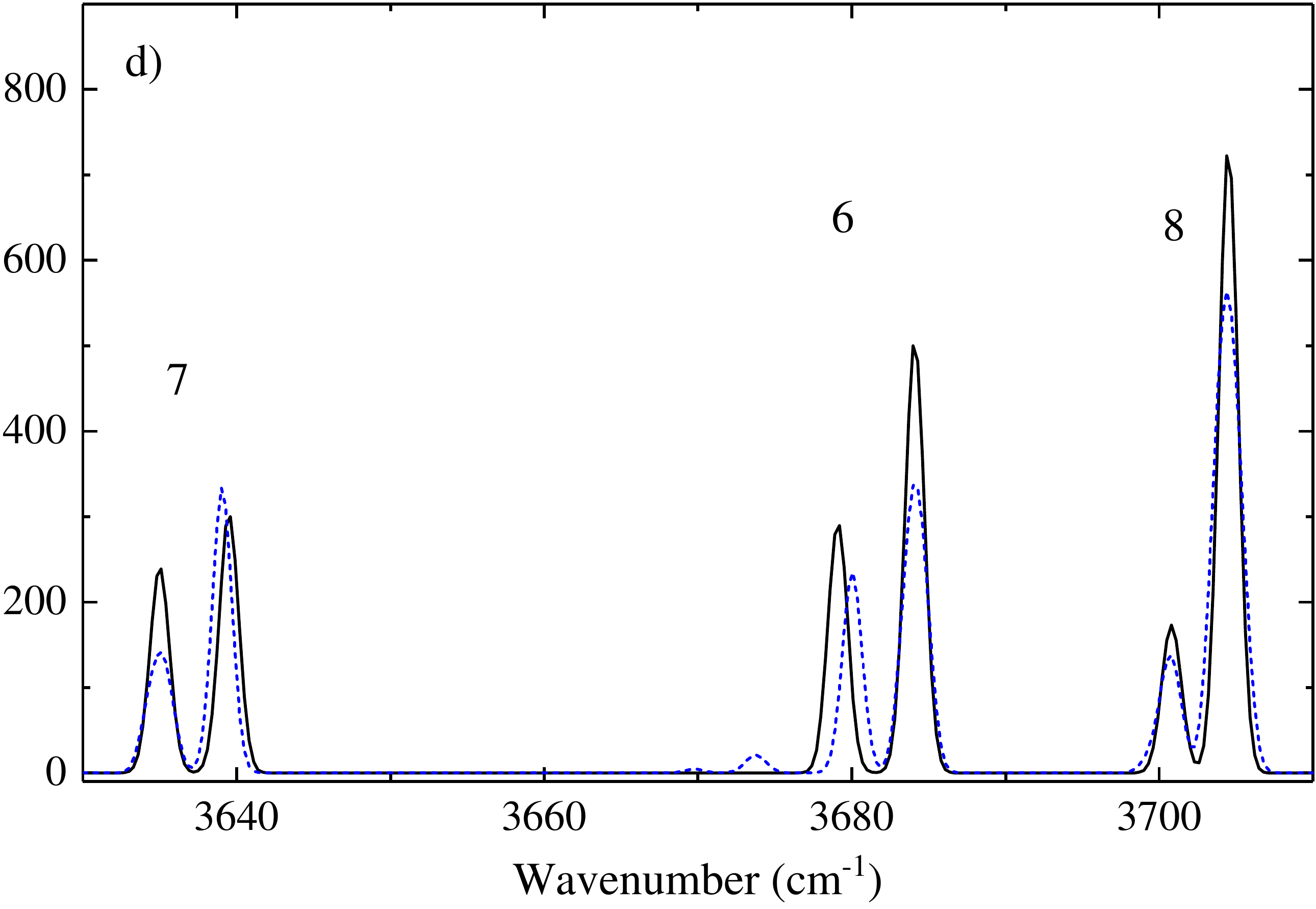}
\caption{\label{fig:Simulations} 
		The synthetic experimental spectra (black solid line) and  the  spectra calculated  with the best fit parameters from  Table~\ref{tab:parameters} (blue dashed line) with filling factor $f=1$, \ortho/ fraction $n_\mathrm{o} =0.7$ and at temperature of 5\,K.
		(a) Rotational transitions in the ground vibrational state, (b) vibrational (1 and 2) and rovibrational (3, 4, 5) transitions of $v_2$  and (c) $v_1$. (d) Rovibrational transitions (6, 7, 8) of $v_3$. }
	\end{figure*}

The synthetic experimental spectra and the best fit spectra are shown in Fig.~\ref{fig:Simulations}, with the best fit parameters given in Table~\ref{tab:parameters}.
The result of the fit overlaps well with the synthetic spectrum,  except for the transition 5 as seen in the first three panels of Fig.~\ref{fig:Simulations}.
While for other transitions one or two Gaussian components were sufficient, the experimental transition lineshape required three components to get a reliable fit of its line area. 
Also, transitions 6 and 7 were represented by two components in the synthetic spectrum, although four peaks are seen in the experimental spectrum, Fig.~\ref{fig:Subl_H2O_3500_5K}.
The additional structure of experimental peaks may originate from the  merohedral disorder as discussed in Section \ref{sec:discussion_crystal_field}. 

Tables~\ref{tab:para_states} and \ref{tab:ortho_states} list energies and the main components of rotational states in the ground vibrational state.
The rotational energies are in qualitative agreement with recent computational estimates~\cite{Felker2020}. 
The $2J+1$ degeneracy of  rotational states is fully removed by the electrostatic field interacting with dipole and quadrupole moment of \water/.

From our fit the permanent dipole moment $\mu^x$ of the encapsulated water  is given  by the absorption cross-section of the IR rotational transitions 3, 4, and 5 in the ground vibrational state.
With the value of $\mu^x$ in hand and by using  the intensities of transitions 1 and 2 we were able to determine the internal static electric field in solid \Csixty/.
The interaction of $\mu^x$ with the crystal electric field  mixes rotational states within ground and excited vibrational states.
For example, in case of \ortho/ water the  components of state $\ket{1_{10}}$ are mixed into  the ground state $\ket{1_{01}}$, Table~\ref{tab:ortho_states}.
This mixing gives the oscillator strength to the  pure vibrational transitions 1 and 2.
As shown in table ~\ref{tab:parameters}, the fitted value of the electric field at the \Csixty/ cage centres is $(110\pm 5)10^6$\,Vm$^{-1}$. 

A splitting of ~ 4\wn/ is observed for  transition 7 and is due to the splitting of the \ortho/ ground state $\ket{1_{01}}$, Fig.~\ref{fig:MolAxes}(c). 
In principle, a splitting could be caused by the interaction of the water electric dipole with an electric field, or by the interaction of the water electric quadrupole moment with an electric field gradient.
The electric field $110\times10^6$\,Vm$^{-1}$ is too small to cause splitting of this magnitude.
This electric field lifts the degeneracy of $m=\pm1$ levels of $\ket{1_{01}}$, but the gap between $m=0$ and $m=\pm1$ levels is due to the quadrupolar interaction.
As the splitting is determined by the  product of  $V_\mathrm{Q} $ and $ Q_{zz}$, it is not possible to have an estimate of how much is the  water quadrupole moment $ Q_{zz}$ screened in \Csixty/.

\begin{table*}
	\begin{ruledtabular}
		\caption{\label{tab:para_states}
			\para/-\wateratsixty/ rotational energies and wavefunctions in the ground vibrational state calculated with the best fit parameters from Table~\ref{tab:parameters}.
			Wavefunction components with amplitude absolute value less than 0.1 are omitted; $\ket{J,-k,-m}\equiv \ket{J,\bar{k},\bar{m}}$. 
		} 
	\begin{tabular}{lld{1.60}}
	$J_{K_aK_c}$		& Energy/\wn/ & \multicolumn{1}{l}{Wavefunction in symmetric top basis \ket{J,k,m}}
	\\\hline

	
	&&\\
	$0_{00}$ &   $0$   &   -0.99 \ket{0, 0, 0}   \\ 
	
	&&\\
	$1_{11}$ &   $33.07$   &   0.49 \big(\ket{1, \bar{1}, \bar{1}} + \ket{1, \bar{1}, 1} - \ket{1, 1, \bar{1}} - \ket{1, 1, 1}\big)   \\ 
	
	&  $ 33.19 $  &   -0.69 \big(\ket{1, \bar{1}, 0} - \ket{1, 1, 0}\big)   \\ 
	
	&  $ 33.75 $  &   0.48 \big(\ket{1, \bar{1}, \bar{1}} - \ket{1, \bar{1}, 1} - \ket{1, 1, \bar{1}} + \ket{1, 1, 1}\big)   \\ 
	
	&&\\
	$2_{02}$ &  $ 67.04 $  &   -0.96 \ket{2, 0, 0} + 0.15 \big(\ket{2, \bar{2}, 0} + \ket{2, 2, 0}\big)   \\ 
	
	&  $ 67.73 $  &   0.68 \big(\ket{2, 0, \bar{1}} + \ket{2, 0, 1}\big)   \\ 
	
	&  $ 68.35 $  &   0.68 \big(\ket{2, 0, \bar{1}} - \ket{2, 0, 1}\big)   \\ 
	
	&  $ 71.27 $  &   0.68 \big(\ket{2, 0, \bar{2}} - \ket{2, 0, 2}\big)   \\ 
	
	&  $ 71.30 $  &   -0.68 \big(\ket{2, 0, \bar{2}} + \ket{2, 0, 2}\big)   \\ 
	
	&&\\
	$2_{11}$&  $ 94.65 $  &   -0.4 \big(\ket{2, \bar{1}, 0} + \ket{2, 1, 0}\big) + 0.41 \big(\ket{2, \bar{1}, \bar{2}} + \ket{2, \bar{1}, 2} + \ket{2, 1, \bar{2}} + \ket{2, 1, 2}\big)   \\ 
	
	&  $ 94.72 $  &   -0.49 \big(\ket{2, \bar{1}, \bar{2}} - \ket{2, \bar{1}, 2} + \ket{2, 1, \bar{2}} - \ket{2, 1, 2}\big)   \\ 
	
	&  $ 94.85 $  &   -0.53 \big(\ket{2, \bar{1}, 0} + \ket{2, 1, 0}\big)  - 0.24 \big(\ket{2, \bar{1}, \bar{2}} + \ket{2, \bar{1}, 2} + \ket{2, 1, \bar{2}} + \ket{2, 1, 2}\big) \\
	&&\multicolumn{1}{l}{$ - 0.21 \big(-\ket{2, \bar{1}, \bar{1}} + \ket{2, \bar{1}, 1} - \ket{2, 1, \bar{1}} + \ket{2, 1, 1}\big)  $} \\ 
	
	&  $ 94.91 $  &  0.45 \big(-\ket{2, \bar{1}, \bar{1}} + \ket{2, \bar{1}, 1} - \ket{2, 1, \bar{1}} + \ket{2, 1, 1}\big) -0.21 \big(\ket{2, \bar{1}, 0} + \ket{2, 1, 0}\big) \\
	&&\multicolumn{1}{l}{$ - 0.15 \big(\ket{2, \bar{1}, \bar{2}} + \ket{2, \bar{1}, 2} + \ket{2, 1, \bar{2}} + \ket{2, 1, 2}\big) $ }\\ 
	
	&  $ 95.01 $  &   0.49 \big(\ket{2, \bar{1}, \bar{1}} + \ket{2, \bar{1}, 1} + \ket{2, 1, \bar{1}} + \ket{2, 1, 1}\big)   \\ 
		\end{tabular}
		
	\end{ruledtabular}
\end{table*}

\begin{table*}
	\begin{ruledtabular}
		\caption{\label{tab:ortho_states}
			\ortho/-\wateratsixty/  rotational energies, zero energy at \para/ ground state $\ket{0_{00}}$, and wavefunctions in the ground vibrational state calculated with the best fit parameters from Table~\ref{tab:parameters}.
			Wavefunction components with amplitude absolute value less than 0.1 are omitted; $\ket{J,-k,-m}\equiv \ket{J,\bar{k},\bar{m}}$. 
		} 
	\begin{tabular}{lld{1.60}}
	$J_{K_aK_c}$		& Energy/\wn/ & \multicolumn{1}{l}{Wavefunction in symmetric top basis \ket{J,k,m}}
	\\\hline


&&\\
$1_{01}$ &  $ 20.89 $  &   0.97 \ket{1, 0, 0}   \\ 

&  $ 24.58 $  &   -0.69 \big(\ket{1, 0, \bar{1}} + \ket{1, 0, 1}\big) + 0.16 \big(\ket{1, \bar{1}, 0} + \ket{1, 1, 0}\big)   \\ 

&  $ 25.51 $  &   -0.7 \big(\ket{1, 0, \bar{1}} - \ket{1, 0, 1}\big)   \\ 

&&\\
$1_{10}$ &  $ 38.88 $  &   0.49 \big(\ket{1, \bar{1}, \bar{1}} - \ket{1, \bar{1}, 1} + \ket{1, 1, \bar{1}} - \ket{1, 1, 1}\big)   \\ 

&  $ 40.03 $  &    0.48 \big(\ket{1, \bar{1}, \bar{1}} + \ket{1, \bar{1}, 1} + \ket{1, 1, \bar{1}} + \ket{1, 1, 1}\big) +0.24 \ket{1, 0, 0}  \\ 

&  $ 43.62 $  &   0.69 \big(\ket{1, \bar{1}, 0} + \ket{1, 1, 0}\big)   \\ 

&&\\
$2_{12}$ &  $ 72.29 $  &   0.71 \big(\ket{2, \bar{1}, 0} - \ket{2, 1, 0}\big)   \\ 

&  $ 73.12 $  &   -0.49 \big(\ket{2, \bar{1}, \bar{1}} + \ket{2, \bar{1}, 1} - \ket{2, 1, \bar{1}} - \ket{2, 1, 1}\big)   \\ 

&  $ 73.36 $  &   -0.49 \big(\ket{2, \bar{1}, \bar{1}} - \ket{2, \bar{1}, 1} - \ket{2, 1, \bar{1}} + \ket{2, 1, 1}\big)   \\ 

&  $ 75.87 $  &   -0.49 \big(\ket{2, \bar{1}, \bar{2}} - \ket{2, \bar{1}, 2} - \ket{2, 1, \bar{2}} + \ket{2, 1, 2}\big)   \\ 

&  $ 75.88 $  &   -0.49 \big(\ket{2, \bar{1}, \bar{2}} + \ket{2, \bar{1}, 2} - \ket{2, 1, \bar{2}} - \ket{2, 1, 2}\big)   \\ 

&&\\
$2_{21}$ &  $ 119.5 $  &   -0.49 \big(\ket{2, \bar{2}, \bar{2}} + \ket{2, \bar{2}, 2} - \ket{2, 2, \bar{2}} - \ket{2, 2, 2}\big)   \\ 

&  $ 119.5 $  &   0.49 \big(\ket{2, \bar{2}, \bar{2}} - \ket{2, \bar{2}, 2} - \ket{2, 2, \bar{2}} + \ket{2, 2, 2}\big)   \\ 

&  $ 122.2 $  &   0.49 \big(\ket{2, \bar{2}, \bar{1}} - \ket{2, \bar{2}, 1} - \ket{2, 2, \bar{1}} + \ket{2, 2, 1}\big)   \\ 

&  $ 122.2 $  &   0.49 \big(\ket{2, \bar{2}, \bar{1}} + \ket{2, \bar{2}, 1} - \ket{2, 2, \bar{1}} - \ket{2, 2, 1}\big)   \\ 

&  $ 123.1 $  &   0.71 \big(\ket{2, \bar{2}, 0} - \ket{2, 2, 0}\big)   \\ 

		\end{tabular}
		
	\end{ruledtabular}
\end{table*}

\section{Discussion \label{sec:discussion}}

\subsection{Vibrations of confined  \water/}

All  eight frequencies of the encapsulated \water/ vibrations found in this work are red-shifted relative to those of free water, see Table~\ref{tab:vib_frequencies}.
The red-shift of the vibrational frequency has been  observed for other endofullerenes, \Htwoatsixty/~\cite{Mamone2009,Ge2011}, HD and \Dtwoatsixty/~\cite{Ge2011D2HD}, and \HFatsixty/~\cite{Krachmalnicoff2016}.
Six water modes have a relative shift between $-2.3\%$ and $-2.5\%$.
Two frequencies, namely the bond-bending mode frequency  $\omega_2$ and  its overtone frequency $2\omega_2$ are shifted by $-1.6\%$ and $-1.5\%$, respectively.

The observed red shifts of the stretching mode frequencies $\omega_1$ and $\omega_3 $ is only partially consistent with previous DFT calculations. 
The DFT-based calculations published by Varadwaj et al.~\cite{Varadwaj2012} do predict vibrational redshifts, while some of the calculations reported by Farimani et al.~\cite{Farimani2013} predict blue shifts rather than red shifts.

The calculation predicts a blue shift of the bending mode $\omega_2$ although ten times less in absolute value than the predicted shift of stretching  modes~\cite{Varadwaj2012}.
The experimental shift of $\omega_2$ is less than that of stretching modes but it is still red-shifted.
The other method, fully coupled nine-dimensional calculation, predicts blue-shifts for all three vibrational modes~\cite{Felker2020}.

\subsection{Translations  of \water/ \label{sec:dissc_trans}}

\begin{table}
	\begin{ruledtabular}
		\caption{\label{tab:translations}
			Translational energies from the ground to the first excited state, $\omega_\mathrm{t}$, of small-molecule endofullerenes and 	the scaling of the harmonic spherical potential $V_2$ for the translational motion of an endohedral  molecule, mass $m_i$,  relative to the \Htwoatsixty/ potential $V_2^{\mathrm{H}_2}$.
			The  anharmonicity is neglected, $\omega_\mathrm{t}\approx \omega^0_\mathrm{t}$, where
		$\omega^0_\mathrm{t}$ is the frequency of an harmonic oscillator, Eq.~\ref{eq:harm_osc_freq}.
		} 
		\begin{tabular}{l|rd{3.1}lr}
		Molecule & $m_i$ &  \omega_\mathrm{t}/\wn/ & $V_2/V_2^{\mathrm{H}_2}$ &Ref.  \\\hline
		\Htwo/ & 2 & 179.5 & 1 &~\cite{Ge2011} \\
			HD & 3 & 157.7 & 1.16 &~\cite{Ge2011D2HD} \\
		\Dtwo/ & 4 & 125.9 & 0.98 &~\cite{Ge2011D2HD} \\
			HF & 20 & 78.6 & 1.92 &~\cite{Krachmalnicoff2016} \\
		\water/ & 18 & 110 & 3.4 &  This work\\
		\water/ & 18 & 162 & 7.3 &Theory\cite{Felker2016,Felker2020}
		\end{tabular}
	\end{ruledtabular}
\end{table}

Table~\ref{tab:translations} lists the measured translational energies from the ground to the first excited state, $\omega_\mathrm{t}$, of small-molecule endofullerenes. 
It is known that the potential of  di-hydrogen in \Csixty/ is anharmonic~\cite{Ge2011,Ge2011D2HD} while the  degree of anharmonicity of \HFatsixty/ and \wateratsixty/ potentials is not known.
For simplicity, we assume that the potential is harmonic for the current case of endohedral molecules, $ \omega_\mathrm{t}\approx \omega^0_\mathrm{t}$, and show  its  scaling relative to \Htwo/ in  Table~\ref{tab:translations}.
In this approximation the harmonic  potential parameter is similar   among the hydrogen isotopologs but  for HF and \water/ $V_2$ is  larger by a factor of 1.9 and 3.4, respectively.
The steeper translational potential for HF and \water/, relative to dihydrogen, is consistent with the larger size of these molecules, and hence their tighter confinement.
The last line of Table~\ref{tab:translations} is the frequency and the harmonic potential of \wateratsixty/ derived by   Bacic and co-workers~\cite{Felker2016,Felker2020} using Lennard-Jones potentials.
The calculated potential is more steep than the experimentally determined potential.

As seen in Fig.~\ref{fig:FIR_spectrum}(b), the absorption line of the \ortho/-\water/  translational mode   is split  by 2.9\wn/.
This splitting may be attributed to the coupling between the endohedral molecule  translation and rotation, associated with the interaction of the non-spherical rotating  molecule with the interior of the \Csixty/ cage, as seen  in \htwoatsixty/~\cite{Mamone2009,Ge2011,Ge2011D2HD}. 
For the particular transition, shown in Fig.~\ref{fig:FIR_spectrum}(b), it is the coupling between translational state with $N=1, L=1$ and rotational state  $\ket{1_{01}}$ of \ortho/ water.
Within  spherical symmetry a good quantum number is $\Lambda=J+L$.
Thus, the translation-rotation coupled translational state $L=1$ and rotational state $J=1$  form three states with $\Lambda$-values 0, 1, and 2.
The calculated energy difference of \ortho/-water $\Lambda=0$ and $\Lambda=1$  states is 8\wn/~\cite{Felker2016} as compared to the experimental value 2.9\wn/.

\subsection{Rotations of \water/ \label{sec:dissc_rot}}

There are two possibilities why the rotational constants of water change  when it is encapsulated. 
First is that the bond length and angles of \water/ change.
The second is that \water/, because of confinement and being non-centrosymmetric, is forced  to rotate about the  ``center of interaction'' which does not  coincide with its nuclear center of mass~\cite{Friedmann1965}.

The rotational constant relates to the moment of inertia $I_{aa}$ as $A=h (8 \pi^2 c_0 I_{aa})^{-1}$ where $c_0$ is the speed of light.
Similarly, $B=h (8 \pi^2 c_0 I_{bb})^{-1}$ for the $b$-axis and $C=h (8 \pi^2 c_0 I_{cc})^{-1}$ for the $c$-axis rotation  ($[A]=$m$^{-1}$ in  SI units and $0.01[A]=$\wn/). 
The moments of inertia are $ I_ {\alpha\alpha}=\sum_i m_i (\beta_i^2 + \gamma_i^2)$, where $\{\alpha_i,\beta_i, \gamma_i\}$ are the Cartesian coordinates of the $i$-th nucleus with mass $m_i$ with  the origin at the nuclear center of mass.

For  non-centrosymmetric molecules, the translation-rotation coupling shifts  the rotational energy levels.
In quantum mechanical terms, the  shift of rotational states  is caused by the   mixing of rotational and translational states by translation-rotation coupling,  example  is  \HDatsixty/~\cite{Xu2008HH_HD_DD,Ge2011D2HD}.
Translation-rotation coupling was not included  in our quantum mechanical model of \wateratsixty/. 
The rotational constants of the model  were free parameters to capture the effect of translation-rotation coupling and the change of the \water/ molecule geometry caused by the \Csixty/ cage.
The rotational constants of the free water are $A_0=27.88$\wn/, $B_0=14.52$\wn/ and  $C_0=9.28$\wn/ in the ground vibrational state~\cite{Toth1991}.
The rotational  constants of endohedral water have relative shifts  $-13\%,   5.5\%,  -8.7\%$ for $A_0, B_0 $ and $C_0$, in the ground vibrational state, Table~\ref{tab:parameters}.
In the following discussion, we use classical arguments to assess whether the shift of rotational levels is caused by the change of water molecule geometry or by the translation-rotation coupling. 

From the symmetry of the \water/ molecule, the nuclear center of mass is on the $b$ axis, Fig.~\ref{fig:MolAxes}.
The shift of \water/ center of rotation in the negative direction of  $b$ axis   decreases   $A$ and $C$ while it does not affect $B$.
 If $B$ changes it must be due to the change of H-O-H bond angle and O-H bond length.
The calculation predicts the lengthening of the H-O bond by 0.0026\AA\,   and decrease of the H-O-H bond angle by $0.87^\circ$ of caged water with respect to free water~\cite{Varadwaj2012}.
With these parameters the relative change of rotational constants $A, B$ and  $C$ is -2.5\%, 0.65\%, -0.46\%, an order of magnitude smaller than derived from the IR spectra of \wateratsixty/.
However, by shifting the rotation center by -0.07\AA\, in the $b$ direction (further away from the oxygen) gives relative changes $-14\%, 0.65\%$ and $-5.2\%$. 
This relative change of $A_{000}$ and $C_{000}$ is not very different from  the values derived from the IR spectra while the relative change of $B_{000}$ is within the error limits, Table~\ref{tab:parameters}.
Thus, it is likely that the dominant contribution to the observed changes of rotational constants in the ground translational state comes not from the change of \water/ molecule bond lengths and bond angle but from the shifting its center of rotation away from the nuclear center of mass of \water/ molecule.

\subsection {Permanent and transition dipole moments}

\begin{table}
	\begin{ruledtabular}
		\caption{\label{tab:dip_moments}
			Absolute values of the dipole moment, unit  D, matrix elements of rotational ($\mu^x$, Eq.\,(\ref{eq:dip00_M_element})) and rovibrational ($\mu^x_{01}, \mu^x_{02}$ and $\mu^z_{03}$, Eq.~(\ref{eq:dip0i_M_element})) transitions  of gaseous \water/, as published,  and of \wateratsixty/ calculated with   Eq.\,(\ref{eq:gas2sixty}) from the rotational and rovibrational absorption cross-sections, Table~\ref{tab:IR_transitions}, and determined by  the fit of IR spectra, Table~\ref{tab:parameters}.
			The cross-section-derived  dipole moments  are the cross-section-error weighted averages of the three  transition, 3, 4 and 5 or 6, 7 and 8 in Table~\ref{tab:IR_transitions}.
		} 
		\begin{tabular}{l|d{1.4}ld{1.13}d{1.15}}
			& \multicolumn{1}{l}{Gas}& Ref. &  \multicolumn{1}{l}{Cross-section} & \multicolumn{1}{l}{ Fit}   \\\hline
			$\mu^x$ & 1.855&\cite{Clough1973}& 0.50\pm 0.05 & 0.474\pm 0.008 \\
			$\mu^x_{01}$ & 0.0153 &\cite{Flaud1975} & (1.23\pm 0.13)10^{-2} & (1.031\pm0.021)10^{-2}\\
			$\mu^x_{02}$ & 0.1269 &\cite{Camy_1976} & (3.38\pm 0.19)10^{-2} & (3.40\pm 0.05)10^{-2}  \\
			$\mu^z_{03}$ &0.0684 &\cite{Flaud1975} & (3.0\pm 0.3)10^{-2} & (2.83\pm 0.07)10^{-2}
		\end{tabular}
		
	\end{ruledtabular}
\end{table}

A comparison of the normalized absorption cross-sections $\langle \sigma_{ji} \rangle$, Eq.~(\ref{eq:norm_crosssection}), for \wateratsixty/    and  for free water is given  in the last column of Table~\ref{tab:IR_transitions}.
In general, for all observed transitions the absorption cross-section of  endohedral water   is smaller than that of free water. 
The ${v_1}$ mode has the smallest relative change while  the largest relative change   is for the combination mode $\omega_1+\omega_3$.

The comparison of \wateratsixty/ and free \water/ absorption cross-sections, Table~\ref{tab:IR_transitions}, enables us to estimate independently from the spectral fit the permanent  and the transition dipole moments of encapsulated \water/,
 \begin{equation}\label{eq:gas2sixty}
 \mu_{ji}^{{\mathrm C}_{60}} = \mu_{ji}^{\mathrm{gas}}\sqrt{\frac{\langle\sigma_{ji}^{{\mathrm C}_{60}}\rangle\,\omega_{ji}^{\mathrm{gas}}}{\langle \sigma_{ji}^{\mathrm{gas}}\rangle\,\omega_{ji}^{{\mathrm C}_{60}}}}.
 \end{equation}
 The results are collected  together with the dipole moments obtained from  the fit of IR spectra in Table~\ref{tab:dip_moments}.
The permanent dipole moment of encapsulated water is nearly four times smaller than of a free \water/.
 The  results of the  IR spectroscopy study are consistent  with the dipole moment determined by the capacitance method, $0.51\pm0.05$\,D~\cite{Meier2015}.
The reduction of fullerene-encapsulated water dipole moment has been predicted by several theoretical calculations~\cite{Ramachandran2005,Yagi2009,Ensing2012,Varadwaj2012}.

\subsection{Effect of solid \Csixty/ crystal field\label{sec:discussion_crystal_field}}

The electric dipolar and quadrupolar interactions of the \wateratsixty/ molecule with the electrostatic field in solid \Csixty/ explains the oscillator strength of the pure $v_1$ and $v_2$ vibrational transitions and the splitting of the rotational states with   $J>0$.

Theoretical work of Felker et al.~\cite{Felker2017} shows that the source of the quadrupole crystal field is the orientation of electron-rich double bonds of 12 nearest-neighbour \Csixty/ relative to the central \wateratsixty/.
When the solid \Csixty/ is cooled below 90\,K a merohedral disorder is frozen in where approximately 85\% of \Csixty/ have electron-rich 6:6 bond (bond between the two hexagonal rings) facing pentagonal ring of a neighbouring cage, the P-orientation~\cite{Heiney1992}.
The rest are H-oriented where the 6:6 bond faces neighbouring cages hexagonal ring.
The calculated   quadrupolar interaction for the P-oriented molecules is  ten times bigger than for the H-oriented molecules~\cite{Felker2017}.
The electric field gradient couples to the   quadrupole moment of  water and  splits the  $\ket{1_{01}}$ rotational state by  $4.2$\wn/, where  the $m=\pm 1$ doublet is above the $m=0$ level~\cite{Felker2017}.
This theoretically predicted splitting of $\ket{1_{01}}$ state for the P-oriented molecules is remarkably close to the observed experimental value, seen as a 4\wn/  splitting of line 7, transition starting from the \ortho/ ground state, Fig.~\ref{fig:Subl_H2O_3500_5K}.
It is not possible to determine the crystal electric  field gradient tensor and the encapsulated water quadrupole moment separately from our  IR spectra.

Further splitting is possible if the   symmetry is lower than $S_6$, but the maximum number of components for $J=1$ remains  three.
However, we see that  line 7 consists of four components instead of three, Fig.\,\ref{fig:Subl_H2O_3500_5K}. 
This suggests that there are two sites with different local electrostatic fields.
Anomalous splitting of triply degenerate phonons into quartets has been seen in solid \Csixty/ by IR spectroscopy~\cite{Homes1994} and was attributed to the merohedral disorder.
Thus, our work and IR study of phonons~\cite{Homes1994} clearly show that there are two different quadrupolar interactions  in solid \Csixty/.
As was proposed by Felker et al. ~\cite{Felker2017}, the crystal field has a different magnitude for  P-oriented sites and for  H-oriented sites. The small population of H-oriented sites (about 15\%)  justifies our spectral fitting with a single quadrupolar interaction.

We assumed that there is an internal electric field in \Csixty/ and this field is a possible reason why the pure vibrational transitions 1 and 2, see Fig.~\ref{fig:MolAxes}(c), become visible in the IR spectrum. 
It is also plausible that 1 and 2 gain intensity through the translation-rotation coupling from the induced dipole moment of translational motion.
However, there is evidence that local electric fields exist in solid \Csixty/ as a result of merohedral disorder~\Csixty/\cite{Alers1992}.
The estimate of un-balanced charge by Alers et al. \cite{Alers1992} was $q=6\times 10^{-3}e$ assuming a dipole moment $\mu= q d_0$,  where $e$ is the electron charge and $d_0=0.7$\,nm is the diameter of a \Csixty/ molecule.
Our estimate is that the electric field $\mathcal{ E}=1.1\times 10^8$\,Vm$^{-1}$ at the center of \Csixty/ cage is created by the  dipole moment with  charge  $q=4.7\times 10^{-3}e$.
These two estimates are very close. 

\Csixty/ has six nearest-neighbor  equatorial cages and three axial cages above and three axial cages below the equatorial plane, following the notation of Ref.~\cite{Felker2017}.
The $z'$ axis of the crystal field coordinate frame is normal to the equatorial plane.
As our fit shows, the electric field is rotated away from the $z'$ axis by $\theta_E\approx 80^\circ$, Table~\ref{tab:parameters}, almost into the equatorial plane.
It is possible that one of the 6 nearest-neighbours in the equatorial plane does not have P-orientation and this   mis-oriented cage is the source of  the electric field.
$\theta_E$ has  a large relative error consistent with the probability to have  the mis-oriented cage in the  equatorial or in the axial position.

\section{Summary \label{sec:summary}}

The infrared absorption spectra of solid \wateratsixty/ samples  were measured close to liquid He temperature and rotational, vibrational, rovibrational, overtone, and combination rovibrational transitions of \water/ were seen. 
The spectral lines were identified  as  \para/ and \ortho/ water transitions  by following the \para/-\ortho/ conversion process over the timescale of hours.
The vibrational frequencies are shifted by -2.4\% relative to free water, except bending mode frequency $\omega_2$ and its overtone $2\omega_2$, where the shift is -1.6\%.
An absorption mode due to the quantized center of mass motion of \water/ in the molecular cage of \Csixty/, was observed at 110\wn/.
The dipole moment of encapsulated water is $0.50\pm 0.05$\,D, approximately 4 times less than for free water and agrees with previous estimates~\cite{Meier2015}.

The rotational and rovibrational spectra were fitted with a quantum mechanical model of a vibrating rotor in electrostatic field with  dipolar and quadrupolar interactions. 
The quadrupolar interaction  splits the $J\geq 1$ rotational states of \water/.
The source of quadrupolar interaction  is the relative orientation of electron-rich chemical bonds relative to pentagonal and hexagonal motifs of \Csixty/ and its 12 nearest neighbours~\cite{Felker2017}.
Further IR study by using pressure to change   the ratio of P- and H-oriented motifs \cite{Sundqvist1999}  would provide more information on the quadrupolar crystal fields of these motifs.
The finite oscillator strength of the fundamental vibrational transitions is attributed to a finite electric field at the centres of the \Csixty/ cages due to merohedral disorder, as has been postulated in different contexts~\cite{Alers1992}. 
Our results are consistent with an internal electric field of $10^8$\,Vm$^{-1}$.
However, it is also plausible that the fundamental vibrational transitions gain intensity through the translation-rotation coupling from the  dipole moment induced by the translational motion, something that can be addressed in further theoretical studies.

To conclude, \water/ in the molecular cavity of \Csixty/ behaves as a vibrating asymmetric top, its dipole moment is reduced, and the translational motion is quantized. 
The splitting of rotational  levels is caused by the quadrupolar interaction with the crystal field of solid \Csixty/. Evidence is found for the existence of a finite electric field at the centres of the \Csixty/ cages in the water-endofullerene, due to merohedral disorder.
 
Two out of three components necessary for a rigorous, comprehensive description of the water translations, rotations and vibrations inside \Csixty/ molecular cage are now in place: firstly, the infrared spectroscopy data reported here and secondly, the  9-dimensional quantum bound-state methodology \cite{Felker2020} plus the theory of symmetry breaking in solid \Csixty/ \cite{Felker2017,Bacic2018FD}. 
What is missing is a high-quality ab initio 9-dimensional  potential energy surface for this system.  

\begin{acknowledgments}
We thank Prof. Zlatko  Ba\v{c}i\'c  for useful discussions.
This research was supported by the Estonian Ministry of Education and Research institutional reseach funding IUT23-3, personal research funding PRG736,  and the European Regional Development Fund project TK134.
We thank EPSRC (UK), grant numbers EP/P009980/1 and EP/T004320/1,  for support.
\end{acknowledgments}

\appendix

\section{Interaction of dipole moment with local electric field\label{sec:E_static}}

The dipole moment in  spherical components is ~\cite{Bunker1998}
\begin{eqnarray}\label{eq:dipSpherCart}
\mu_{+1} &=& -\frac{1}{\sqrt 2 }(\mu_{x}+i \mu_{y}),\nonumber\\
\mu_{-1} &=& \frac{1}{\sqrt 2 }(\mu_{x}- i \mu_{y}),\nonumber\\
\mu_0 &=& \mu_z.
\end{eqnarray}

The dipole moment of water in the Cartesian molecule coordinate frame, as shown in Fig.~\ref{fig:MolAxes}, is 
$ {\bm \mu}^{M}= \{-\mu_x,0,0\} $, 
where we use a convention that the dipole moment is directed from the negative charge to the positive charge.
Then,  from Eq.~\ref{eq:dipSpherCart} the dipole moment in spherical components is
\begin{equation}\label{eq:water_dip_moment_mol_spher_Appendix}
\{\mu^{M}_m\}=\frac{\mu_x}{\sqrt{2}}\{-1,0, 1\}, \quad m=-1, 0, +1.
\end{equation}

We consider the coupling of \water/ dipole moment to the local electric field, $\vect{\mathcal{E}}$, with spherical coordinates $\{\mathcal{E},\phi_E, \theta_E\}$ in the crystal frame $C$ (frame where the electric field gradient  tensor is defined) and in the coordinate  frame $E$ of electric field  $\{E^E_m\}=\mathcal{E} \{0,1,0\}$, i.e. along $z_E$ axis.
Corresponding Euler angles are $\Omega_{C\rightarrow E}=\{\phi_E, \theta_E,0\}$ and 
$\Omega_{E\rightarrow C}=\{\pi,\theta_E,-\pi-\phi_E \}$.
The dipole moment of the molecule in the frame $E$ is 
\begin{eqnarray}
\mu^{E}_{m} & = & \sum_{m'=-1}^{1} \left[ \WignerD{1}{mm'}{\Omega_{E\rightarrow C}} \right]^* \mu^{C}_{m'}.  \end{eqnarray}
Here we used  Wigner $D$-functions  relating  the components of a spherical rank $j$ irreducible tensor $T_{jm}$ in coordinate frames $A$ and $B$ \cite{Zare1988}:
\begin{eqnarray}
T^{B}_{jm} & = & \sum_{m'=-j}^{j}  \WignerD{j}{m'm}{\Omega_{A\rightarrow B}}T^{A}_{jm'}
\end{eqnarray}
and
\begin{eqnarray}\label{eq:rank_J_transform}
T^{A}_{jm}&=&\sum_{m'=-j}^{j} \left[ \WignerD{j}{mm'}{\Omega_{A\rightarrow B}} \right]^* T^{B}_{jm'},  
\end{eqnarray}
where $\Omega_{A\rightarrow B}=\{ \phi,\theta,\chi\} $  are Euler angles transforming coordinate frame $A$ into frame $B$.
The angles for the inverse transformation are $\Omega_{B\rightarrow A}=\{ \pi-\chi,\theta,-\pi-\phi\} $~\cite{Varshalovich1988}.

The interaction of molecular dipole moment $\mu^{M}_{m}$ with electric field is 
\begin{eqnarray}\label{eq:Hed_appendix}
\mathcal{H}_\mathrm{ed}&=&-\sum_{m=-1}^1 (-1)^m\mathcal{E}^{E}_{-m} \mu^{E}_m \nonumber\\
&=&- \sum_{m=-1}^1  (-1)^m \mathcal{E}^{E}_{-m} \left[ \sum_{m'=-1}^{1} \left[ \WignerD{1}{mm'}{\Omega_{E\rightarrow C}} \right]^* \mu^{C}_{m'}\right] \nonumber\\
&=&- \sum_{m,m'=-1}^1  (-1)^m \mathcal{E}^{E}_{-m} \left[ \WignerD{1}{mm'}{\Omega_{E\rightarrow C}} \right]^*  \mu^{C}_{m'} \nonumber\\
&=&- \sum_{m,m'=-1}^1  (-1)^m \mathcal{E}^{E}_{-m} \left[ \WignerD{1}{mm'}{\Omega_{E\rightarrow C}} \right]^* \nonumber\\
&&\qquad \qquad \times \left[\sum_{m^{\prime\prime}=-1}^1  \left[ \WignerD{1}{m'm^{\prime\prime}}{\Omega_{C\rightarrow M}} \right]^*\mu^{M}_{m^{\prime\prime}} \right]\nonumber.
\end{eqnarray}
The minus sign in front of sum in the  formula above  is  consistent with $ \mathcal{H}_\mathrm{ed}= - \mathcal{ E}\cdot { \mu}$ if the  spherical components of vectors $\mathcal{ E}$ and $\mu$  are defined as in (\ref{eq:dipSpherCart}).

\section{Interaction of dipole moment with  electric field of radiation\label{sec:E_radiation}}

Here we derive the electric dipole transition matrix elements of a molecule,  part (\ref{eq:dip_moment_aver}) of absorption cross section (\ref{eq:cross-section_op}).
The derivation of (\ref{eq:dip_moment_aver}) starts from 
\begin{equation}\label{eq:H_ed_squared}
S_{fi} = \frac{1}{e^2} |\bra{f} \mathcal{H}_{\mathrm{ed}} \ket{i}|^2,
\end{equation}
and 
\begin{equation}\label{eq:H_ed}
\mathcal{H}_{\mathrm{ed}} =  -\sum^1_{m=-1}(-1)^m e^{A}_{-m} \mu^{A}_m ,
\end{equation}
where the electric field of radiation,  
$\vect{e}=\{e^A_m\}$ and the dipole moment of a molecule, $\mu^{A}_m$, are in the space-fixed frame $A$; $e^2=\sum_{m=-1}^{1}(e_m^A)^2 $ .

The Euler angles of transformation of crystal frame $C$ to space-fixed frame $A$ and vice versa are $\Omega_{C\rightarrow A}=\{\phi_R, \theta_R,0\}$ and 
$\Omega_{A\rightarrow C}=\{\pi,\theta_R,-\pi-\phi_R \}$.

The dipole moment in  frame $A$ is
\begin{eqnarray}
\mu^{A}_{m} & = & \sum_{m'=-1}^{1} \left[ \WignerD{1}{mm'}{\Omega_{A\rightarrow C}} \right]^* \mu^{C}_{m'},  
\end{eqnarray}
where $\mu^{C}_{m}$ is the dipole moment in the crystal frame $C$.

The absolute value of matrix element squared is 
\begin{eqnarray}
S_{fi}&=& \nonumber\\
&=& e^{-2} \left|\bra{j}\sum_{m=-1}^1  (-1)^m e^{A}_{-m}\mu^{A}_{m} \ket{i}\right|^2\nonumber\\
&=& e^{-2} \left|\sum_{m,m'=-1}^1  (-1)^m e^{A}_{-m}\left[\WignerD{1}{mm'}{\Omega_{A\rightarrow C}} \right]^*\bra{j}\mu^{C}_{m'} \ket{i}\right|^2\nonumber\\
&=& e^{-2} \sum_{m_1,m_1',m_2,m_2'=-1}^1  (-1)^{m_1+m_2} e^{A}_{-m_1} \left[e^{A}_{-m_2}\right]^*\nonumber\\
&&\hspace{12pt}\times\left[\WignerD{1}{m_1m'_1}{\Omega_{A\rightarrow C}} \right]^*\WignerD{1}{m_2m'_2}{\Omega_{A\rightarrow C}}\nonumber\\
&&\hspace{12pt}\times\bra{j}\mu^{C}_{m'_1} \ket{i}\bra{j}\mu^{C}_{m'_2} \ket{i}^*. \label{eq:el_dipole_abs_squared}	
\end{eqnarray}

\subsection{Random orientation of crystals}

We assume all the crystals are identical and there are no  other static fields outside the crystal what could brake the directional isotropy.
We average (\ref{eq:el_dipole_abs_squared}) over the random orientation of crystal coordinate frames with respect to the space-fixed frame, $\Omega_{A\rightarrow C}$, and get:
\begin{eqnarray}
\left<S_{fi}\right>_{\Omega_{A\rightarrow C}}&=&\nonumber\\
&=&\frac{1}{3e^2}\sum_{m,m'=-1}^1  \left|e^{A}_{-m}\right|^2 \left|\bra{j}\mu^{C}_{m'} \ket{i}\right|^2 \nonumber\\
&=&\frac{1}{3e^2}\left(\sum_{m =-1}^1  \left|e^{A}_{-m}\right|^2 \right)\left(\sum_{m' =-1}^1\left|\bra{j}\mu^{C}_{m'} \ket{i}\right|^2\right) \nonumber\\
&=&\frac{1}{3}\left(\sum_{m' =-1}^1\left|\bra{j}\mu^{C}_{m'} \ket{i}\right|^2\right). \label{eq:el_dipole_abs_squared_averaged}
\end{eqnarray}
where we used a  property of rotation matrices~\cite{Zare1988},
\begin{eqnarray}
&&\int \left[\WignerD{j_1}{m_1m'_1}{\Omega} \right]^*\WignerD{j_2}{m_2m'_2}{\Omega}\mathrm{d}\Omega\\
& &\hspace{12pt} = \frac{8\pi^2}{2j_1+1} \delta_{j_1j_2} \delta_{m_1m_2}\delta_{m'_1m'_2},\nonumber
\end{eqnarray}
and 
\begin{equation}
\int_0^\pi \sin \theta_R \mathrm{d}\theta_R \int_0^{2\pi}\mathrm{d}\phi_R\int_0^{2\pi}\mathrm{d}\chi_R= 8\pi^2.
\end{equation}
If the sample is in the powder form then it follows from  Eq.~(\ref{eq:el_dipole_abs_squared_averaged}) that the absorption is independent of light polarization.

\subsection{Transition  matrix element and separation of coordinates \label{sec:transition_matrix_element}}

The absorption of radiation by a molecule, Eq.~(\ref{eq:cross-section_op}),  depends on the matrix elements of an electric dipole moment between the initial and final states, 
\begin{equation}\label{eq:eldip_matrix_elem_internal}
\sum_{\sigma =-1}^1  \left|\bra{\Phi'}\mu^{A}_{\sigma} \ket{\Phi}\right|^2,
\end{equation}
where $\mu^{A}_{\sigma} $ is the molecule diopole moment  in the space-fixed coordinate frame.
Molecule wavefunction consists of nuclear spin wavefunction $\ket{Im_I}$, electron $\ket{\Phi^{e}}$ and electron spin wavefunction $\ket{Sm_S}$, vibration wavefunction $\ket{\Phi^{v}}$, and rotation wavefunction $\ket{\Phi^{r}}$~\cite{Bunker1998}:
\begin{equation}\label{eq:mol_wave_fn}
\ket{\Phi}=\ket{Im_I}\ket{Sm_S}\ket{\Phi^{e}\Phi^{v}\Phi^{r}}.
\end{equation}

Using these separable wavefunctions the matrix element (\ref{eq:eldip_matrix_elem_internal}) is~\cite{Bunker1998}:
\begin{eqnarray}\label{eq:eldip_matrix_elem_separable}
&&\bra{\Phi'}\mu^{A}_{\sigma} \ket{\Phi}=\left< I'm'_I \right|\left. Im_I \right>\left<  S'm'_S \right|\left. Sm_S  \right>\\
&& \hspace{12pt}\times \sum_{\sigma'=-1}^{1} \bra{\Phi^{r'}} \WignerD{1}{\sigma\sigma'}{\Omega_{A\rightarrow M}}^* \ket{\Phi^{r}} \bra{\Phi^{v'}}     \mu_{\sigma'}^{M(e)}\ket{\Phi^{v}},\nonumber
\end{eqnarray}
where 
\begin{equation}\label{eq:eldip_matrix_elem_2}
\mu_{\sigma}^{M(\mathrm{e})}=\bra{\Phi^{e'}}     \mu_{\sigma}^{M}\ket{\Phi^{e}}
\end{equation}
and we have taken into account that the electric dipole moment  does not depend   on nuclear and electron spin coordinates.
Furthermore, if the energies of initial and final states of the transition are independent of spin projections $m_I$ and $m_s$, the summation over initial and final states in the transition probability leads to degeneracy factors $g_I=2I+1$ and $g_S=2S+1$ in (\ref{eq:cross-section_op}).
The electron spin is zero in the ground electronic state of \water/, thus $g_S=1$. 
Degeneracy of \para/ molecules ($I=0$) is $g_I^{(\mathrm p)}=1$ and \ortho/  molecules ($I=1$) is $g_I^{(\mathrm o)}=3$.

 If the  electronic orbital  does not change in the transition then (\ref{eq:eldip_matrix_elem_internal}) is the  molecule electric dipole moment in the ground electronic state $\ket{\Phi^{e}}$,
\begin{equation}\label{eq:eldip_matrix_elem_internal_diagonal}
\mu_{\sigma}^{M(\mathrm{eg})}\equiv \bra{\Phi^{e}}     \mu_{\sigma}^{M}\ket{\Phi^{e}}.
\end{equation}
For the rest of the discussion we use a shorthand notation $\mu_{\sigma}^{M}$ for the molecule dipole moment in the ground electronic state, $\mu_{\sigma}^{M(\mathrm{eg})}$.

Using Cartesian coordinates the dipole moment in the ground vibrational state is
\begin{equation}\label{eq:dip00_M_element}
\mu^x=\bra{000}\mu_{x}^{M}\ket{000}.
\end{equation}
In the quantum mechanical model of \wateratsixty/ what we used to fit the IR spectra we set the dipole moment  equal to $\mu^x$ in three  excited vibrational states $\ket{100}$, $\ket{010}$, and $\ket{001}$.

The vibrational transition dipole moments are
\begin{eqnarray}\label{eq:dip0i_M_element}
\mu^x_{01}=\bra{100}\mu_{x}^{M}\ket{000},\nonumber\\
\mu^x_{02}=\bra{010}\mu_{x}^{M}\ket{000},\\
\mu^z_{03}=\bra{001}\mu_{z}^{M}\ket{000}.\nonumber
\end{eqnarray}
The relation between spherical and Cartesian dipole moment components is given by Eq.~\ref{eq:dipSpherCart}.

\section{Fitting of synthetic spectra with quantum mechanical model and model parameter error estimation\label{sec:fittingANDerror} }

We determined the Hamiltonian parameters and the dipole moments, parameter set $\parvec/=\{a_1,  \ldots, a_\nu, \ldots ,a_M \}$, by finding  the paramater set $\parvecmin/$ what gives the minimum value, $\chisqmin/$,  of function
\begin{equation}\label{eq:chisq_sum}
\chisq/ = \sum_{i=1}^{N} \left[ S(\nu_i) - f(\nu_i;\parvec/) \right]^2
\end{equation}
where $S(\nu_i)$ is the synthetic spectrum, argument frequency $\nu_i$, generated from the fit of the experimental spectra, and $ f(\nu_i;\parvec/)$ is the spectrum calculated from the model with $M$ parameters $\parvec/=\{a_1,  \ldots, a_\nu, \ldots ,a_M \}$;
$N$ is the number of points in the spectrum.
The goal is to minimize  $\chisq/$ over parameters  $\parvec/$.
The result is $\chisq/_{\mathrm{min}}$ and $\parvec/_{\mathrm{min}}$.

Lets define matrix
\begin{equation}\label{eq:F_element}
F_{i\nu} = \frac{\partial f(\nu_i;\parvec/)}{\partial a_\nu}
\end{equation}
and dispersion  matrix $\matx{V}$, Eq.\,12 in~\cite{Albritton1976},
\begin{equation}\label{eq:R_matrix}
V_{\mu\nu}=\sum_{i=1}^{N}F^T_{\mu i}F_{i\nu}. 
\end{equation} 	
The covariance matrix, Eq.\,11 in~\cite{Albritton1976},
\begin{equation}\label{eq:M_matrix}
\matx{\Theta}=\sigma^2 \matx{V}^{-1},
\end{equation} 
where $\sigma^2=\chisq/_{min}/(N-M)$ and $\matx{V}^{-1}$ is inverse matrix of $\matx{V}$, $\matx{V}^{-1}\matx{V}=\matx{1}$. 

The estimated variance of  parameter $a_\nu$ is  
\begin{equation}\label{eq:a_error}
\Delta a_\nu = \sqrt{\Theta_{\nu\nu}}=\sqrt{\frac{\chisq/_{\mathrm{min}}(\matx{V}^{-1})_{\nu\nu}}{N-M}}.
\end{equation} 

The correlation matrix $\matx{C}$ is
\begin{equation}\label{eq:CorrelationMatrix}
C_{\nu\mu}= \frac{\Theta_{\nu\mu}}{\sqrt{\Theta_{\nu\nu}\Theta_{\mu\mu}}}.
\end{equation}

The element of $\matx{F}$, Eq.(\ref{eq:F_element}), is 
\begin{equation}\label{eq:F_element_numeric}
F_{i\nu} = \frac{f(\nu_i;\parvec/_{\mathrm{min}}+\delta a_\nu )-f(\nu_i;\parvec/_{\mathrm{min}})}{\delta a_\nu},
\end{equation}
where $\parvec/_{min}$ minimizes $\chisq/$ and $\delta a_\nu $ is a small variation of parameter $ a_\nu$.

We change the sum  for an integral over $\nu$ in (\ref{eq:chisq_sum}),
\begin{equation}\label{eq:chisq_int}
\chisq/ =\sum_{k} \int_{\nu_{k1}}^{\nu_{k2}} A_k^2 \left[ S(\nu) - f(\nu;\parvec/) \right]^2\mathrm{d}\nu,
\end{equation}
where the experimental spectrum  is available in several spectral ranges $\{ \nu_{k1},\nu_{k2} \}$ indexed by $k$.
$A_k$ is the weight factor for each spectral range $k$.
$A_k$ is chosen so that   the strongest lines  for each range $k$ are equal. 
The  spectrum $S(\nu)$ is calculated with constant linewidth using line areas and frequencies from the fits of the experimental spectra.

By inserting  (\ref{eq:F_element_numeric}) into  (\ref{eq:R_matrix}) and using continuum limit we get  matrix $\matx{V}$,
\begin{eqnarray} \label{eq:V_matrix_contiinum}
V_{\mu\nu} & = & \frac{1}{\delta a_\mu \delta a_\nu} \sum_{k} A_k^2 \int_{\nu_{k1}}^{\nu_{k2}}  \mathrm{d}\nu \nonumber\\ 
& &\left[ f(\nu;\parvec/_{\mathrm{min}}+\delta a_\mu ) \; f(\nu;\parvec/_{\mathrm{min}}+\delta a_\nu )\right. \nonumber\\
&&\hspace{12pt} - f(\nu;\parvec/_{\mathrm{min}}+\delta a_\mu )\; f(\nu;\parvec/_{\mathrm{min}})\nonumber\\
&&\hspace{12pt}-  f(\nu;\parvec/_{\mathrm{min}})\; f(\nu;\parvec/_{\mathrm{min}}+\delta a_\nu )\nonumber\\
&&\hspace{12pt} + \left. f(\nu;\parvec/_{\mathrm{min}})^2\right].
\end{eqnarray}

We define the number of experimental points $N$ as the number of  lines fitted in the experimental spectra multiplied by two as each line has two parameters, area and frequency.
We estimate the parameter error with Eq.\,(\ref{eq:a_error}) using (\ref{eq:chisq_int}) and (\ref{eq:V_matrix_contiinum}).



%

\end{document}